\documentclass[aps, prl, reprint, superscriptaddress]{revtex4-2}

\usepackage{newtxtext, newtxmath}
\usepackage{graphicx}
\usepackage{braket}
\usepackage{dcolumn}
\usepackage{bm}
\usepackage{soul}
\usepackage{hyperref}
\usepackage{color}
\definecolor{bblue}{rgb}{0.8, 0.0, 0}
\hypersetup{
    colorlinks=true,        % false: boxed links; true: colored links
    linkcolor=bblue,         % color of internal links
    citecolor=bblue,        % color of links to bibliography
    filecolor=magenta,      % color of file links
    urlcolor=bblue           % color of external links
}

\begin{document}

\title{Signature of spin-triplet exciton condensations in LaCoO$_{3}$ at ultrahigh magnetic fields up to 600 T}

\author{Akihiko~Ikeda}
\email[Corresponding author: ]{a-ikeda@uec.ac.jp}
\affiliation{Institute for Solid State Physics, University of Tokyo, Kashiwa, Chiba 277-8581, Japan}
\affiliation{Department of Engineering Science, University of Electro-Communications, Chofu, Tokyo 182-8585, Japan}
\author{Yasuhiro~H.~Matsuda}
\affiliation{Institute for Solid State Physics, University of Tokyo, Kashiwa, Chiba 277-8581, Japan}
\author{Keisuke~Sato}
\affiliation{National Institute of Technology, Ibaraki College, Hitachinaka, Ibaraki 312-0011, Japan}
\author{Yuto~Ishii}
\author{Hironobu~Sawabe}
\author{Daisuke~Nakamura}
\email[Present address: RIKEN Center for Emergent Matter Science (CEMS), Wako 351-0198, Japan]{}
\author{Shojiro~Takeyama}
\affiliation{Institute for Solid State Physics, University of Tokyo, Kashiwa, Chiba 277-8581, Japan}
\author{Joji~Nasu}
\affiliation{Department of Physics, Tohoku University, Sendai, Miyagi 980-8578, Japan}
\affiliation{PRESTO, Japan Science and Technology Agency, Honcho Kawaguchi, Saitama 332-0012, Japan}

\date{\today}

\begin{abstract}
\paragraph*{\rm{\textbf{Abstract}}}
Bose-Einstein condensation of electron-hole pairs, exciton condensation, has been effortfully investigated since predicted 60 years ago. 
Irrefutable evidence has still been lacking due to experimental difficulties in verifying the condensation of the charge neutral and non-magnetic spin-singlet excitons.
Whilst, condensation of spin-triplet excitons is a promising frontier because spin supercurrent and spin-Seebeck effects will be observable.
A canonical cobaltite LaCoO$_{3}$ under very high magnetic fields is a propitious candidate, yet to be verified.
Here, we unveil the exotic phase diagram of LaCoO$_{3}$ up to 600 T generated using the electromagnetic flux compression method and the state-of-the-art magnetostriction gauge.
We found the continuous magnetostriction curves and a bending structure, which suggest the emergence of two distinct spin-triplet exciton condensates.
By constructing a phenomenological model, we showed that quantum fluctuations of excitons are crucial for the field-induced successive transitions.
The spin-triplet exciton condensation in a cobaltite, which is three-dimensional and thermally equilibrated, opens up a novel venue for spintronics technologies with spin-supercurrent such as a spin Josephson junction.
\end{abstract}

\maketitle

 \begin{figure}[]
 \includegraphics[width = \linewidth]{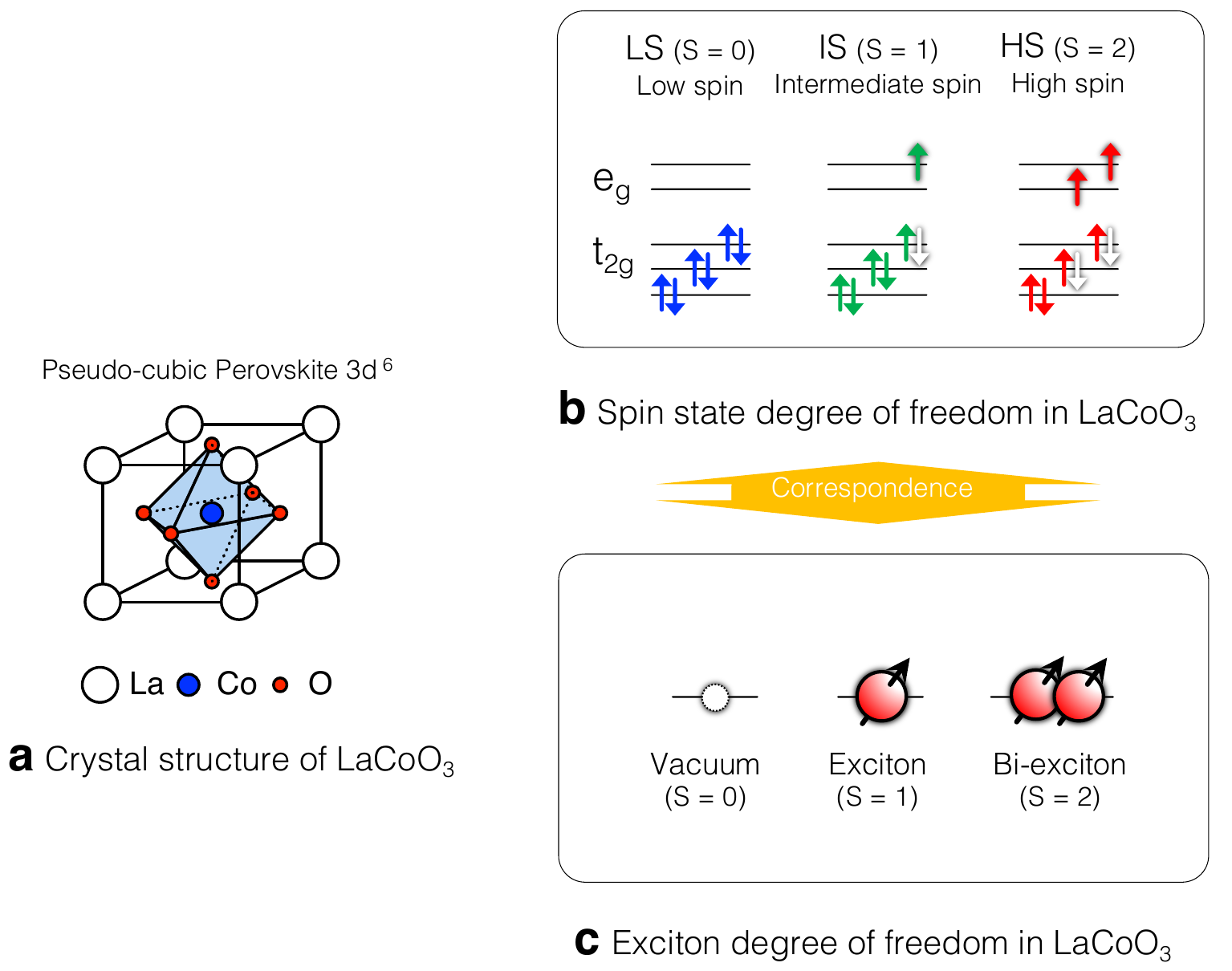}%
 \caption{\textbf{Schematic illustrations describing the correspondence between spin-states and exciton states in LaCoO$_{3}$.}
\textbf{a} Schematic crystal structure of LaCoO$_{3}$.
\textbf{b} Schematic drawing of spin-state degrees of freedom of LaCoO$_{3}$.
\textbf{c} Schematic drawing of excitonic degrees of freedom in LaCoO$_{3}$.
LS and a pair of shaded electron and hole in (\textbf{b}) correspond to vacuum and an exciton in (\textbf{c}), respectively.
\label{intro}}
 \end{figure}

 \begin{figure}[bh!]
 \includegraphics[width = \linewidth]{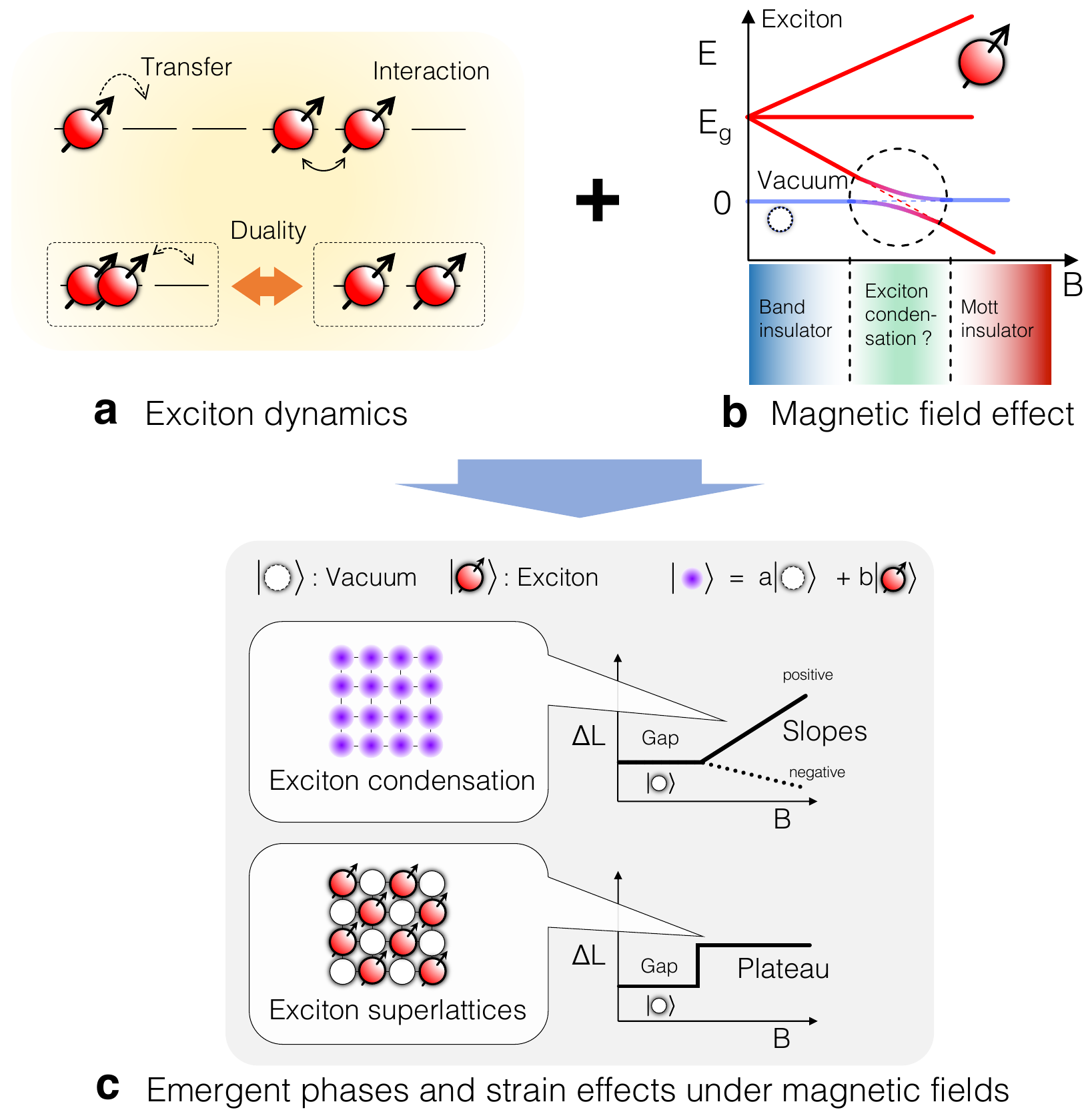}%
\caption{\textbf{Schematic illustrations describing the possible occurrence of exotic orderings of excitons in LaCoO$_{3}$ under high magnetic fields.}
\textbf{a} Schematic drawing of the interactions and dynamics of the excitons.
\textbf{b} Schematic drawing of the magnetic field effect on the exciton states in LaCoO$_{3}$. 
\textbf{c} Expected emergence of the magnetic field induced phases in  LaCoO$_{3}$, and the resultant magnetostriction curves. 
 \label{intro2}}
 \end{figure}

\paragraph*{\rm{\textbf{Introduction}}}
In condensed matters, novel phases emerge at boundaries of incompatible phases, where competing interactions and fluctuations trigger spontaneous symmetry breakings.
Insulator state via the Bose-Einstein condensation of electron-hole pairs, excitonic insulator or exciton condensation, has been eagerly investigated over 50 years since its prediction \cite{Halperin}, which emerges at the boundary region between semi-metals and semi-conductors and their photo-excited states.
However, compelling evidence has been lacking, due to the difficulty in verifying the condensation of charge neutral and spin-singlet excitons \cite{KogarScience2017, LuNC2017, LaiNat2007, YanyuNP2021, BretscherSA2021}.

On the other hand, the condensation of the spin-triplet excitons is more beneficial to investigate thanks to observable magnetic phenomena such as spin-supercurrent \cite{YuanSA2018, JiangPRL2020} and spin Seebeck effect \cite{NasuPRB2021}, which are attractive for potential applications in spintronics and quantum computing technologies.
Condensation of spin-triplet bosons is of fundamental interest in analogy with the spin-triplet superconductivity \cite{Manago2019, Ran2019} and superfluidity of $^{3}$He \cite{HalperinARCMP2019}.
The condensation of spin-triplet excitons has been rarely considered in conventional studies of exciton condensations \cite{KogarScience2017, LuNC2017, LaiNat2007, YanyuNP2021, BretscherSA2021}.

Recently, theories have predicted that spin-triplet exciton condensation is stable in perovskite cobaltites with strongly correlated electrons \cite{KanekoPRB2015, KunesJPCM2015, NasuPRB2016}.
This is based on the fact that the spin-state degrees of freedom are inherent in the cobaltites [see Fig. \ref{intro}a and \ref{intro}b] due to the competing Hund's coupling and the crystal field splitting, which can also be viewed as the novel degrees of freedom of atomic size excitons [see Fig. \ref{intro}c].
The spin-triplet exciton condensation is essentially the quantum hybridization of the distinct exciton states.
The hybridization is due to the delocalization, interaction, and duality of excitons [see Fig. \ref{intro2}a] that are promoted when the energies of the vacuum and exciton states are in proximity by such external fields as magnetic fields [See Fig. \ref{intro2}b].
The LS and HS states correspond to the Mott insulator and the band insulator phases, respectively \cite{KanekoPRB2015}.
Thereby, the spin-triplet exciton condensation occurs at the boundary region between the incompatible phases [See Fig. \ref{intro2}b].
%Note that it is also a multipole order of spin and orbital degrees of freedom \cite{KanekoPRB2016}.

 \begin{figure*}
 \includegraphics[width = \linewidth]{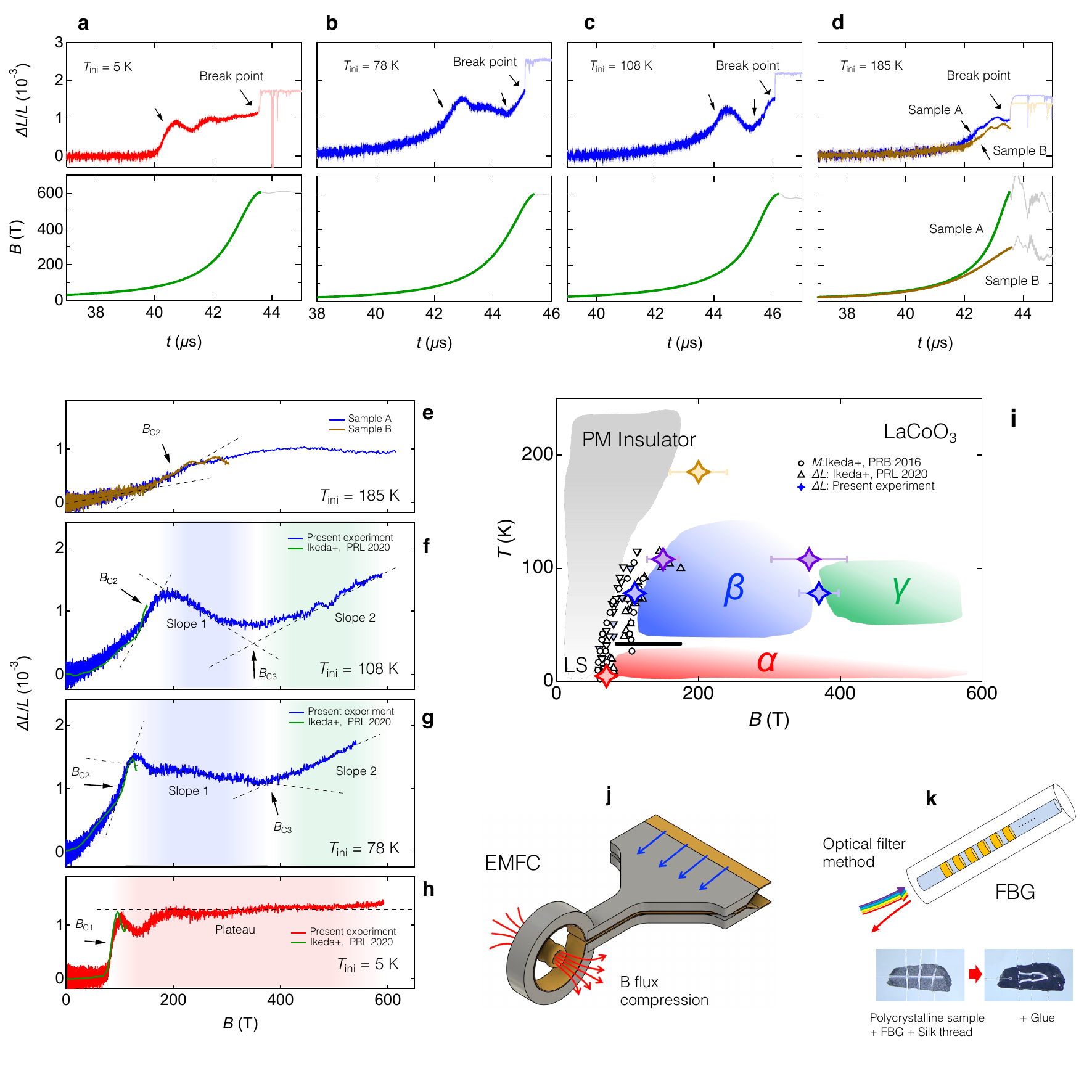}%
 \caption{\textbf{Magnetostriction data up to 600 T for LaCoO$_{3}$ at various temperatures and the summarizing phase diagram.}
 \textbf{a-d} Result of magnetostriction measurement of LaCoO$_{3}$ shown as a function of time, along with the magnetic field profiles generated using the EMFC method.
 The initial sample temperatures $T_{\rm{ini}}$ are varied from 5 K to 185 K.
 \textbf{e-h} Magnetostriction data of LaCoO$_{3}$ shown as a function of magnetic field at various $T_{\rm{ini}}$, along with the reference data (Green curves) previously obtained using STC  \cite{IkedaPRL2020}.
 \textbf{i} Summary of the magnetostriction measurement of LaCoO$_{3}$ on a magnetic field-temperature plane up to 600 T along with the previously obtained data using magnetization \cite{IkedaLCO2016} and magnetostriction measurements \cite{IkedaPRL2020}.
Sample A and Sample B in (\textbf{d}) and (\textbf{e}) represent the data of LaCoO$_{3}$ samples positioned at the center (sample A) and 7 mm off the center (sample B) of the coil in the axial direction.
 \textbf{j} An illustration of the main coil and liner during the EMFC.
 \textbf{k} Illustrations of the FBG fiber and sample glued to the FBG fiber used in the experiment.
 \label{result}}
 \end{figure*}

A doped perovskite cobaltite Pr$_{0.5}$Ca$_{0.5}$CoO$_{3}$ and its family compounds are the first candidate for the spin-triplet exciton condensation, exhibiting an insulating and paramagnetic ground state at low temperatures \cite{HejtmanekEPJB2013, KunesPRB2014}.
The origin of the insulator phase is, however, not completely uncovered because the significant valence transition of Pr complicates the phase transition.
Next, the most well-known cobaltite LaCoO$_{3}$ has been claimed to be a promising candidate when placed under very high magnetic fields exceeding 100 T \cite{SotnikovSR2016, TatsunoJPSJ2016}.
LaCoO$_{3}$ is an archetypal compound having a variety of spin-states, such as low-spin (LS: $S=0$, $t_{2g}^{6}e_{g}^{0}$), intermediate spin (IS: $S=1$, $t_{2g}^{5}e_{g}^{1}$) and high spin (IS: $S=2$, $t_{2g}^{4}e_{g}^{2}$) states [See Fig. \ref{intro}b], which are viewed as vacuum, an exciton, and bi-exciton states, respectively [See Fig. \ref{intro}c].
The spin-triplet exciton condensation will emerge when the magnetic field eliminates the spin-gap and changes the vacuum state (LS) to a magnetic exciton state (IS) or bi-exciton state (HS) [See Fig. \ref{intro2}b].

However, the experimental investigation has been challenging due to the need for very high magnetic fields beyond 100 T which is necessitated by the large spin-gap of $\sim 100$ K.
Besides, one needs to probe the exciton states at such high magnetic fields.
As solutions to these difficulties, recently, we reported the generation of 1200 T with electromagnetic flux compression (EMFC) method using a newly installed capacitor banks system in ISSP \cite{NakamuraRSI2018}.
Furthermore, we succeeded in implementing a state-of-the-art high-speed 100 MHz strain gauge using fiber Bragg grating (FBG) and optical filter method \cite{AIkedaFBG2017}, which have enabled us to measure magnetostriction in the $\mu$s-pulsed 1000 T environment.
Magnetostriction is a crucial probe of the exciton states because the density of excitons is coupled to the lattice volume of LaCoO$_{3}$, where exciton and bi-exciton states have larger ionic volumes with the occupation of $e_{g}$ orbitals that is spatially more extended than $t_{2g}$ orbitals as can be seen in the correspondence of spin-states and exciton states [See Figs. \ref{intro}b and \ref{intro}c.]
Solidifications and Bose-Einstein condensations of excitons result in plateaux and slope of exciton density \cite{SotnikovSR2016, TatsunoJPSJ2016, SotnikovPRB2017}.
Thus, we expect that they also result in plateaux and slope in magnetostriction curves at very low temperatures, which is schematically shown in Fig. \ref{intro2}c.
Note that the correspondences are analogs to the magnetization plateaux and slopes in magnon solids and superfluids, respectively \cite{RiceScience2002, ZapfRMP2014}.

Previously in LaCoO$_{3}$, the field-induced phase transitions are uncovered below 30 K ($\alpha$ phase) \cite{SatoJPSJ2009, MoazPRL2012, RotterSR2014} and above 30 K ($\beta$ phase), which extend beyond 100 T, using magnetization and magnetostriction measurements up to 200 T generated by single turn coil (STC) method \cite{IkedaLCO2016, IkedaPRL2020}.
The two phases are considered superlattice formation ({\it i.e.} solidification) of the excitons, bi-excitons, and vacuum based on the observation of plateaux of magnetostrictions.
Considering that the magnetization below 200 T is far from the full polarization of the spin-states, further orderings of excitons are expected to appear above 200 T.
So far, only one magnetization study up to 500 T \cite{Platonov2012} has been reported.
Now, we are fully equipped for the investigation of the evolution of the excitonic states in LaCoO$_{3}$ beyond 200 T.

 \begin{figure}[]
 \includegraphics[width = 0.9\linewidth]{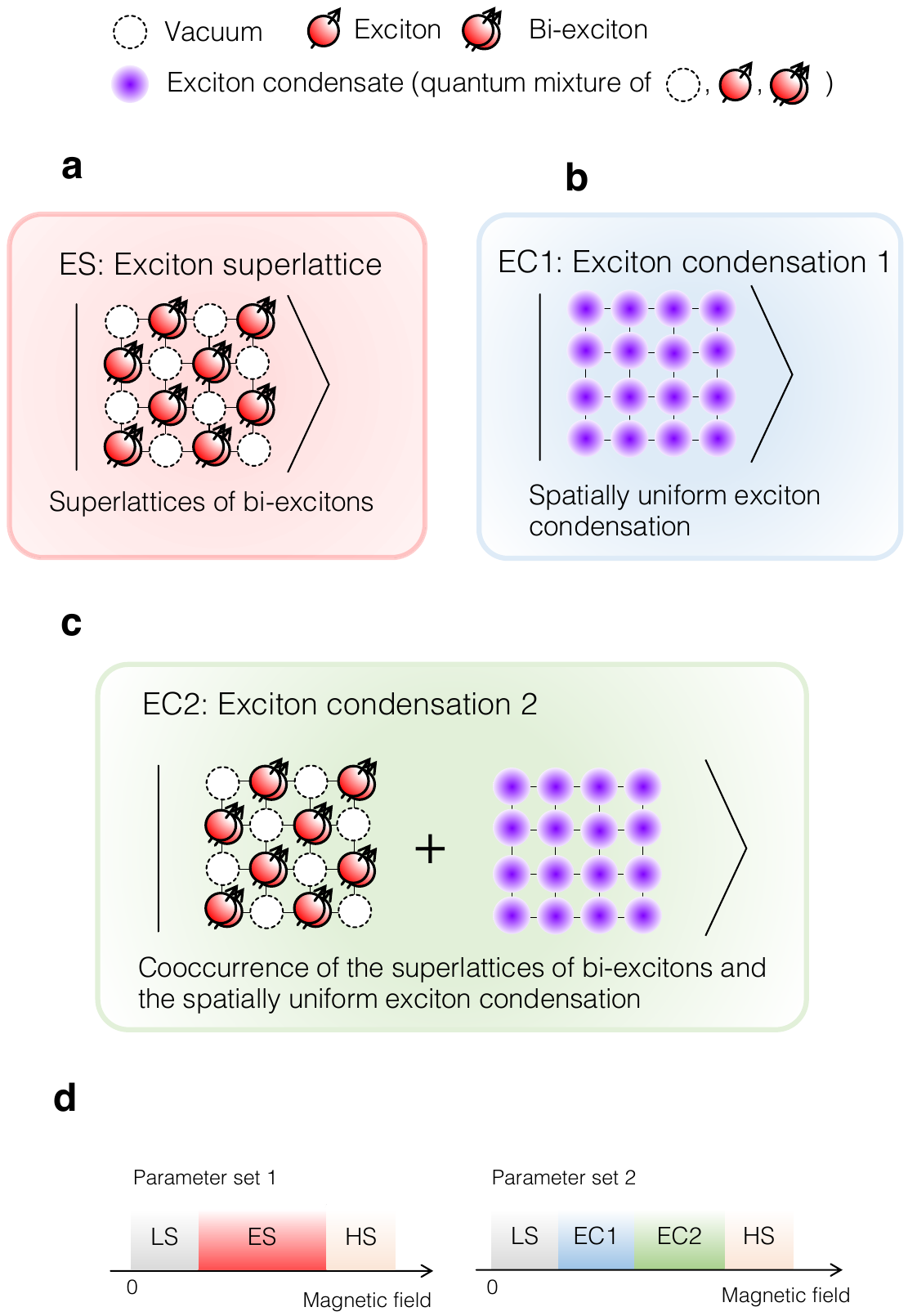}%
 \caption{
\textbf{Schematic illustrations of calculated excitonic phases.}
\textbf{a} Exciton superlattice (ES), where exciton and vacuum form a superlattice. 
\textbf{b} Exciton condensation 1 (EC1), where Bose-Einstein condensation of excitons occur spatially uniformly.
\textbf{c} Exciton condensation 2 (EC2), where Bose-Einstein condensation of excitons occur with a translational symmetry breaking.
\textbf{d} Magnetic field dependence of the mean-field phases with the assumption of distinct parameter sets.
See Supplementary Note 2 for the details of the parameters.
 \label{nasu}
 }
 \end{figure}

\paragraph*{\rm{\textbf{Results}}}
Figures \ref{result}a-\ref{result}h show the results of magnetostriction measurements of polycrystalline LaCoO$_{3}$ up to 600 T with the initial sample temperatures $T_{\rm{ini}} =  5$, 78, 108, and 185 K.
Figures \ref{result}a-\ref{result}d show the magnetostriction data and magnetic fields as a function of time.
Magnetostriction data as a function of magnetic fields are shown in Figs. \ref{result}e-\ref{result}h.
The magnetostriction data up to 150 T in Ref. \cite{IkedaPRL2020} are shown for comparison using green curves in Figs. \ref{result}f-\ref{result}h.
The transition points are indicated in Figs. \ref{result}e-\ref{result}h and are summarized using star-shaped symbols in Fig. \ref{result}i along with the previous results up to 150 T \cite{IkedaLCO2016, IkedaPRL2020} shown using filled and open triangles and circles.
The transitions and features below 200 T well reproduce the previous observations up to 150 T [See Figs. \ref{result}e-\ref{result}h] where the transition fields are denoted as $B_{\rm{C1}}$ and $B_{\rm{C2}}$.
 The propagation of a shock wave inside the sample initiated by the strong transitions of $B_{\rm{C1}}$ and $B_{\rm{C2}}$ are present as indicated in Figs. \ref{result}a-\ref{result}d, which is also reported in Refs. \cite{IkedaPRL2020, RicoPNAS}.
We distinguish the intrinsic features and the shock propagation by carefully observing the sound speed and temperature dependence.
See Supplementary Note 1 for the details.
The temperature changes of the sample during the application of the $\mu$-second pulsed magnetic field are considered significant at $T_{\rm{ini}} = 5$ K, while it is negligible at $T_{\rm{ini}} > 30$ K based on the sufficiently large heat capacity above 30 K \cite{KyomenPRB2003}.% and preliminary measurement of magnetocaloric effects up to 65 T \cite{Kihara}.

The magnetostriction plateau of the $\alpha$ phase at $T_{\rm{ini}} = 5$ K that appears above 70 T is found to persist up to 600 T, revealing its significant stability under magnetic fields.
On the other hand, the $\beta$ phase is found to show a negative slope of magnetostriction [See slope 1 in Fig. \ref{result}g] between 170 T and 380 T.
Furthermore, we find a bending structure indicating a new phase transition at 380 T denoted as $B_{\rm{C3}}$ from the $\beta$ phase to a novel state in the data at $T_{\rm{ini}} =  78$ K.
The new state is termed the $\gamma$ phase, which is characterized by the positive slope beyond 380 T [See slope 2 in Fig. \ref{result}g].
The sharp slopes of slope 1 and slope 2 at $T_{\rm{ini}} = 78$ K become smeared in the data obtained at $T_{\rm{ini}} = 108$ K as shown in Fig. \ref{result}f.
Thus, the sharp slopes of slope 1 and slope 2 obtained at $T_{\rm{ini}} =  78$ K are not thermal origins.
But rather, we argue that they originate in quantum fluctuations of excitons such as exciton condensations as schematically shown in Fig. \ref{intro2}c.
Previously in Ref. \cite{IkedaPRL2020}, the $\beta$ phase was falsely considered a plateau due to the measurement regions limited below 200 T.
Furthermore, the negative slope itself disappears at the $T_{\rm{ini}} = 185$ K data as shown in Fig. \ref{result}e, while the feature of the transition of $B_{\rm{C2}}$ remains.
These features indicate that both $\beta$ and $\gamma$ phases become unstable at $\sim$100 K, that the $\gamma$ phase is completely absent at $\sim$185 K, and also that the feature of the $\beta$ phase does not completely disappear at $\sim$185 K.

\paragraph*{\rm{\textbf{Discussion}}}
Here, we discuss the origins of the $\alpha$, the $\beta$, and the $\gamma$ phases in LaCoO$_{3}$ induced at ultrahigh magnetic fields.
We claim that the $\alpha$ phase is a superlattice of spin-triplet excitons and vacuum, based on the present observation of the magnetostriction plateau at $T_{\rm{ini}} = 5$ K shown in Fig. \ref{result}h and that the magnetization of $\alpha$ phase is only 1/8 or 1/4 of the saturation of the exciton states \cite{SatoJPSJ2009, MoazPRL2012}.
In the exciton superlattice, the number of the exciton is discrete \cite{NasuPRB2016}, leading to the magnetostriction plateau as is shown in the data at 5 K in Fig. \ref{result}h.
This is in good agreement with the previous observations and considerations on the $\alpha$ phase \cite{SatoJPSJ2009, MoazPRL2012, RotterSR2014, IkedaPRL2020}.
The remarkable stability of the $\alpha$ phase that sustains up to 600 T suggests the localized nature of the spin-triplet excitons in the $\alpha$ phase of LaCoO$_{3}$.

On the other hand, the $\beta$ and the $\gamma$ phases show magnetostriction slopes denoted as slope 1 and slope 2 in Figs. \ref{result}f and \ref{result}g.
Based on the observation that they are nonthermal in origin, they indicate that exciton density varies continuously as a function of magnetic fields.
Considering the firm coupling of lattice volume and exciton density in LaCoO$_{3}$, these behaviors imply the continuous change of lattice volume as a function of magnetic fields, which is a characteristic of Bose-Einstein condensation of excitons \cite{SotnikovSR2016, TatsunoJPSJ2016, SotnikovPRB2017}.
In a condensate of excitons, the single site wave function of Co is described by the linear combination of $\ket{\phi_{\rm{EC}}} = a\ket{0}+b\ket{1}+c\ket{2}$, where $\ket{0}$, $\ket{1}$, $\ket{2}$ represents the vacuum, an exciton, and a bi-exciton state, respectively. A global wave-function can be expressed as $\ket{\psi_{\rm{EC}}} = \prod_{i}\ket{\phi_{\rm{EC}}}_{i}$.
The continuous change of magnetostriction and magnetization is realized because the coefficients $a$, $b$, $c$ can change continuously in exciton condensates as a function of magnetic field  \cite{TatsunoJPSJ2016, SotnikovSR2016}.
Another possible origin of continuous magnetostriction is thermal activation.
We rule out this possibility based on the negative magnetostriction observed in slope 1 of Fig. \ref{result}g.
No negative thermal expansion is reported in LaCoO$_{3}$ \cite{Radaelli}.
The possibilities of orbital order and domain re-orientations are raised based on the facts that LaCoO$_{3}$ has an orbital degree of freedom in IS and HS states and that it is a slightly anisotropic pseudo-cubic system \cite{NoguchiPRB2002}.
We rule out these possibilities because we used polycrystalline samples.

It is well known that the exciton condensations are difficult to be evidenced by a single experiment.
One needs various macroscopic and microscopic experiments \cite{KunesJPCM2015}.
Here, we tentatively construct a phenomenological model Hamiltonian just for a qualitative discussion of the possible appearance of multiple exciton condensations in LaCoO$_{3}$ at ultrahigh magnetic fields.
We start from the five-orbital Hubbard model with six electrons per site for the electronic states of Co ions.
In the strong correlation limit, there are contributions from the exchange processes of $e_{g}$ and $t_{2g}$ electrons between neighboring Co ions.
This leads to the itinerancy of excitons (${\cal H}_{\rm{trans}}$) and the fusion of two excitons to a bi-exciton (${\cal H}_{\rm{dual}}$).
The model introduced here consists of these two quantum processes in addition to the local exciton energy with the Zeeman field (${\cal H}_{\rm{loc}}$) and classical interaction between neighboring (bi-)excitons (${\cal H}_{\rm{int}}$).
Note that the present model includes the bi-exciton state, which is in contrast to the previous literature on exciton condensation, where only the exciton state is taken into account in Refs. \cite{SotnikovSR2016, TatsunoJPSJ2016}.
Note also that the model is a quantum upgrade of the classical model introduced in Ref. \cite{MoazPRL2012}.
The details are given in Supplementary Note 2.

Here, we just show the essence of the calculated results in Figs. \ref{nasu}a-d, for simplicity.
When the contributions from the quantum processes are weak, the phase transition occurs from the vacuum ground state to an exciton superlattice (ES) as depicted in Fig. \ref{nasu}a with a plateau of exciton density.
In the ES phase, the exciton state does not change with increasing magnetic fields, which would result in a plateau of magnetostriction.
ES is a possible candidate for the $\alpha$ phase.
Further, with sufficient contribution from the duality term, the calculated results show that the phase transitions occur from the vacuum ground state to two kinds of exciton condensations with increasing magnetic fields.
As schematically shown in Fig. \ref{nasu}b, the exciton condensation 1 (EC1) is a spatially uniform exciton condensation made of a mixture of all the exciton states as $\ket{\phi_{\rm{EC1}}} = a\ket{0}+b\ket{1}+c\ket{2}$.
EC1 is followed by the appearance of exciton condensation 2 (EC2) with increasing magnetic fields.
In EC2, both order parameters of exciton superlattice and exciton condensation become finite simultaneously as schematically shown in Fig. \ref{nasu}c.
In EC1 and EC2, the exciton states change continuously with increasing magnetic fields, which would result in slopes of magnetostriction.
EC1 and EC2 are possible candidates for the $\beta$ and $\gamma$ phases.
As shown in Fig. \ref{nasu}d, EC1 and EC2 appear with a different parameter set from that showed ES.
The large lattice change between the $\alpha$ and $\beta$ phase \cite{IkedaPRL2020} can be an origin of such variations of parameter sets in LaCoO$_{3}$ at above 100 T.

Lastly, we discuss the result in light of the spin-crossover physics in LaCoO$_{3}$.
We argue that the HS-IS duality and the competing parameters are of crucial importance when considering the controversial properties of LaCoO$_3$.
It has been a long-standing controversy for over half a century how to understand the peculiar temperature evolution of magnetic and electric properties of LaCoO$_{3}$ in terms of the spin-states.
The involvement of various spin-states LS, IS, and HS, and also their interactions have been proposed \cite{SCObook}.
The controversy has been long-lasting largely due to the fact that no long-range order has been found in the temperature evolution of LaCoO$_{3}$  \cite{SCObook}.
The present study and previous papers  \cite{IkedaPRL2020, MoazPRL2012}  have reported that various ordered phases emerge under ultrahigh magnetic fields in LaCoO$_{3}$.
However, previously reported models can not account for the presently observed complex phase diagram (Fig. \ref{result}i) and especially the continuous change of the magnetostriction in the $\beta$ and the $\gamma$ phases. 
This indicates that the HS-IS duality,  which is included in the present model, is an important factor in LaCoO$_{3}$.
Further microscopic experiments such as x-ray diffraction \cite{IkedaAPL2022} and thorough theoretical investigations are mandatory to clarify this point.

\paragraph*{\rm{\textbf{Methods}}}
The magnetic fields up to 600 T are generated using the electromagnetic flux compression method \cite{NakamuraRSI2018} in the Institute for Solid State Physics, the University of Tokyo, Japan.
The magnetostriction measurement is performed using our state-of-the-art high-speed strain gauge utilizing fiber Bragg grating and optical filter method \cite{AIkedaFBG2017}.
Polycrystalline samples of LaCoO$_{3}$ were prepared by the solid-state-reaction of predried La$_{2}$O$_{3}$ and CoO at 1300 C$^{\circ}$ in air.
The sample is cut out in $1 \times 3 \times 0.5$ mm.
FBG fiber is glued tightly onto a specimen of the sample using Stycast 1266.

\paragraph*{\rm{\textbf{Data availability}}}
The data that support the findings of this study are available in Zenodo with the identifier [DOI: 10.5281/zenodo.7627257] \cite{IkedaZENODO2023}.

\paragraph*{\rm{\textbf{References}}}
\bibliography{lco}

\begin{thebibliography}{37}%
\makeatletter
\providecommand \@ifxundefined [1]{%
 \@ifx{#1\undefined}
}%
\providecommand \@ifnum [1]{%
 \ifnum #1\expandafter \@firstoftwo
 \else \expandafter \@secondoftwo
 \fi
}%
\providecommand \@ifx [1]{%
 \ifx #1\expandafter \@firstoftwo
 \else \expandafter \@secondoftwo
 \fi
}%
\providecommand \natexlab [1]{#1}%
\providecommand \enquote  [1]{``#1''}%
\providecommand \bibnamefont  [1]{#1}%
\providecommand \bibfnamefont [1]{#1}%
\providecommand \citenamefont [1]{#1}%
\providecommand \href@noop [0]{\@secondoftwo}%
\providecommand \href [0]{\begingroup \@sanitize@url \@href}%
\providecommand \@href[1]{\@@startlink{#1}\@@href}%
\providecommand \@@href[1]{\endgroup#1\@@endlink}%
\providecommand \@sanitize@url [0]{\catcode `\\12\catcode `\$12\catcode
  `\&12\catcode `\#12\catcode `\^12\catcode `\_12\catcode `\%12\relax}%
\providecommand \@@startlink[1]{}%
\providecommand \@@endlink[0]{}%
\providecommand \url  [0]{\begingroup\@sanitize@url \@url }%
\providecommand \@url [1]{\endgroup\@href {#1}{\urlprefix }}%
\providecommand \urlprefix  [0]{URL }%
\providecommand \Eprint [0]{\href }%
\providecommand \doibase [0]{https://doi.org/}%
\providecommand \selectlanguage [0]{\@gobble}%
\providecommand \bibinfo  [0]{\@secondoftwo}%
\providecommand \bibfield  [0]{\@secondoftwo}%
\providecommand \translation [1]{[#1]}%
\providecommand \BibitemOpen [0]{}%
\providecommand \bibitemStop [0]{}%
\providecommand \bibitemNoStop [0]{.\EOS\space}%
\providecommand \EOS [0]{\spacefactor3000\relax}%
\providecommand \BibitemShut  [1]{\csname bibitem#1\endcsname}%
\let\auto@bib@innerbib\@empty
%</preamble>
\bibitem [{\citenamefont {Halperin}\ and\ \citenamefont
  {Rice}(1968)}]{Halperin}%
  \BibitemOpen
  \bibfield  {author} {\bibinfo {author} {\bibfnamefont {B.~I.}\ \bibnamefont
  {Halperin}}\ and\ \bibinfo {author} {\bibfnamefont {T.~M.}\ \bibnamefont
  {Rice}},\ }\bibfield  {title} {\bibinfo {title} {Possible anomalies at a
  semimetal-semiconductor transistion},\ }\href
  {https://doi.org/10.1103/RevModPhys.40.755} {\bibfield  {journal} {\bibinfo
  {journal} {Rev. Mod. Phys.}\ }\textbf {\bibinfo {volume} {40}},\ \bibinfo
  {pages} {755} (\bibinfo {year} {1968})}\BibitemShut {NoStop}%
\bibitem [{\citenamefont {Kogar}\ \emph {et~al.}(2017)\citenamefont {Kogar},
  \citenamefont {Rak}, \citenamefont {Vig}, \citenamefont {Husain},
  \citenamefont {Flicker}, \citenamefont {Joe}, \citenamefont {Venema},
  \citenamefont {MacDougall}, \citenamefont {Chiang}, \citenamefont {Fradkin},
  \citenamefont {van Wezel},\ and\ \citenamefont
  {Abbamonte}}]{KogarScience2017}%
  \BibitemOpen
  \bibfield  {author} {\bibinfo {author} {\bibfnamefont {A.}~\bibnamefont
  {Kogar}}, \bibinfo {author} {\bibfnamefont {M.~S.}\ \bibnamefont {Rak}},
  \bibinfo {author} {\bibfnamefont {S.}~\bibnamefont {Vig}}, \bibinfo {author}
  {\bibfnamefont {A.~A.}\ \bibnamefont {Husain}}, \bibinfo {author}
  {\bibfnamefont {F.}~\bibnamefont {Flicker}}, \bibinfo {author} {\bibfnamefont
  {Y.~I.}\ \bibnamefont {Joe}}, \bibinfo {author} {\bibfnamefont
  {L.}~\bibnamefont {Venema}}, \bibinfo {author} {\bibfnamefont {G.~J.}\
  \bibnamefont {MacDougall}}, \bibinfo {author} {\bibfnamefont {T.~C.}\
  \bibnamefont {Chiang}}, \bibinfo {author} {\bibfnamefont {E.}~\bibnamefont
  {Fradkin}}, \bibinfo {author} {\bibfnamefont {J.}~\bibnamefont {van Wezel}},\
  and\ \bibinfo {author} {\bibfnamefont {P.}~\bibnamefont {Abbamonte}},\
  }\bibfield  {title} {\bibinfo {title} {Signatures of exciton condensation in
  a transition metal dichalcogenide},\ }\href
  {https://doi.org/10.1126/science.aam6432} {\bibfield  {journal} {\bibinfo
  {journal} {Science}\ }\textbf {\bibinfo {volume} {358}},\ \bibinfo {pages}
  {1314} (\bibinfo {year} {2017})}\BibitemShut {NoStop}%
\bibitem [{\citenamefont {Lu}\ \emph {et~al.}(2017)\citenamefont {Lu},
  \citenamefont {Kono}, \citenamefont {Larkin}, \citenamefont {Rost},
  \citenamefont {Takayama}, \citenamefont {Boris}, \citenamefont {Keimer},\
  and\ \citenamefont {Takagi}}]{LuNC2017}%
  \BibitemOpen
  \bibfield  {author} {\bibinfo {author} {\bibfnamefont {Y.~F.}\ \bibnamefont
  {Lu}}, \bibinfo {author} {\bibfnamefont {H.}~\bibnamefont {Kono}}, \bibinfo
  {author} {\bibfnamefont {T.~I.}\ \bibnamefont {Larkin}}, \bibinfo {author}
  {\bibfnamefont {A.~W.}\ \bibnamefont {Rost}}, \bibinfo {author}
  {\bibfnamefont {T.}~\bibnamefont {Takayama}}, \bibinfo {author}
  {\bibfnamefont {A.~V.}\ \bibnamefont {Boris}}, \bibinfo {author}
  {\bibfnamefont {B.}~\bibnamefont {Keimer}},\ and\ \bibinfo {author}
  {\bibfnamefont {H.}~\bibnamefont {Takagi}},\ }\bibfield  {title} {\bibinfo
  {title} {\rm{Zero-gap semiconductor to excitonic insulator transition in
  Ta}$_{2}$\rm{NiSe}$_{5}$},\ }\href {https://doi.org/10.1038/ncomms14408}
  {\bibfield  {journal} {\bibinfo  {journal} {Nat. Commun.}\ }\textbf {\bibinfo
  {volume} {8}},\ \bibinfo {pages} {14408} (\bibinfo {year}
  {2017})}\BibitemShut {NoStop}%
\bibitem [{\citenamefont {Lai}\ \emph {et~al.}(2007)\citenamefont {Lai},
  \citenamefont {Kim}, \citenamefont {Utsunomiya}, \citenamefont {Roumpos},
  \citenamefont {Deng}, \citenamefont {Fraser}, \citenamefont {Byrnes},
  \citenamefont {Recher}, \citenamefont {Kumada}, \citenamefont {Fujisawa},\
  and\ \citenamefont {Yamamoto}}]{LaiNat2007}%
  \BibitemOpen
  \bibfield  {author} {\bibinfo {author} {\bibfnamefont {C.~W.}\ \bibnamefont
  {Lai}}, \bibinfo {author} {\bibfnamefont {N.~Y.}\ \bibnamefont {Kim}},
  \bibinfo {author} {\bibfnamefont {S.}~\bibnamefont {Utsunomiya}}, \bibinfo
  {author} {\bibfnamefont {G.}~\bibnamefont {Roumpos}}, \bibinfo {author}
  {\bibfnamefont {H.}~\bibnamefont {Deng}}, \bibinfo {author} {\bibfnamefont
  {M.~D.}\ \bibnamefont {Fraser}}, \bibinfo {author} {\bibfnamefont
  {T.}~\bibnamefont {Byrnes}}, \bibinfo {author} {\bibfnamefont
  {P.}~\bibnamefont {Recher}}, \bibinfo {author} {\bibfnamefont
  {N.}~\bibnamefont {Kumada}}, \bibinfo {author} {\bibfnamefont
  {T.}~\bibnamefont {Fujisawa}},\ and\ \bibinfo {author} {\bibfnamefont
  {Y.}~\bibnamefont {Yamamoto}},\ }\bibfield  {title} {\bibinfo {title}
  {\rm{Coherent zero-state and $\pi$-state in an exciton-polariton condensate
  array}},\ }\href {https://doi.org/10.1038/nature06334} {\bibfield  {journal}
  {\bibinfo  {journal} {Nature}\ }\textbf {\bibinfo {volume} {450}},\ \bibinfo
  {pages} {529} (\bibinfo {year} {2007})}\BibitemShut {NoStop}%
\bibitem [{\citenamefont {Jia}\ \emph {et~al.}(2021)\citenamefont {Jia},
  \citenamefont {Wang}, \citenamefont {Chiu}, \citenamefont {Song},
  \citenamefont {Yu}, \citenamefont {J\''{a}ck}, \citenamefont {Lei},
  \citenamefont {Klemenz}, \citenamefont {Cevallos}, \citenamefont {Onyszczak},
  \citenamefont {Fishchenko}, \citenamefont {Liu}, \citenamefont {Farahi},
  \citenamefont {Xie}, \citenamefont {Xu}, \citenamefont {Watanabe},
  \citenamefont {Taniguchi}, \citenamefont {Bernevig}, \citenamefont {Cava},
  \citenamefont {Schoop}, \citenamefont {Yazdani},\ and\ \citenamefont
  {Wu}}]{YanyuNP2021}%
  \BibitemOpen
  \bibfield  {author} {\bibinfo {author} {\bibfnamefont {Y.}~\bibnamefont
  {Jia}}, \bibinfo {author} {\bibfnamefont {P.}~\bibnamefont {Wang}}, \bibinfo
  {author} {\bibfnamefont {C.-L.}\ \bibnamefont {Chiu}}, \bibinfo {author}
  {\bibfnamefont {Z.}~\bibnamefont {Song}}, \bibinfo {author} {\bibfnamefont
  {G.}~\bibnamefont {Yu}}, \bibinfo {author} {\bibfnamefont {B.}~\bibnamefont
  {J\''{a}ck}}, \bibinfo {author} {\bibfnamefont {S.}~\bibnamefont {Lei}},
  \bibinfo {author} {\bibfnamefont {S.}~\bibnamefont {Klemenz}}, \bibinfo
  {author} {\bibfnamefont {F.~A.}\ \bibnamefont {Cevallos}}, \bibinfo {author}
  {\bibfnamefont {M.}~\bibnamefont {Onyszczak}}, \bibinfo {author}
  {\bibfnamefont {N.}~\bibnamefont {Fishchenko}}, \bibinfo {author}
  {\bibfnamefont {X.}~\bibnamefont {Liu}}, \bibinfo {author} {\bibfnamefont
  {G.}~\bibnamefont {Farahi}}, \bibinfo {author} {\bibfnamefont
  {F.}~\bibnamefont {Xie}}, \bibinfo {author} {\bibfnamefont {Y.}~\bibnamefont
  {Xu}}, \bibinfo {author} {\bibfnamefont {K.}~\bibnamefont {Watanabe}},
  \bibinfo {author} {\bibfnamefont {T.}~\bibnamefont {Taniguchi}}, \bibinfo
  {author} {\bibfnamefont {B.~A.}\ \bibnamefont {Bernevig}}, \bibinfo {author}
  {\bibfnamefont {R.~J.}\ \bibnamefont {Cava}}, \bibinfo {author}
  {\bibfnamefont {L.~M.}\ \bibnamefont {Schoop}}, \bibinfo {author}
  {\bibfnamefont {A.}~\bibnamefont {Yazdani}},\ and\ \bibinfo {author}
  {\bibfnamefont {S.}~\bibnamefont {Wu}},\ }\bibfield  {title} {\bibinfo
  {title} {\rm{Evidence for a monolayer excitonic insulator}},\ }\href
  {https://doi.org/10.1038/s41567-021-01422-w} {\bibfield  {journal} {\bibinfo
  {journal} {Nat. Phys.}\ }\textbf {\bibinfo {volume} {18}},\ \bibinfo {pages}
  {87} (\bibinfo {year} {2021})}\BibitemShut {NoStop}%
\bibitem [{\citenamefont {Bretscher}\ \emph {et~al.}(2021)\citenamefont
  {Bretscher}, \citenamefont {Andrich}, \citenamefont {Murakami}, \citenamefont
  {Golez}, \citenamefont {Remez}, \citenamefont {Telang}, \citenamefont
  {Singh}, \citenamefont {Harnagea}, \citenamefont {Cooper}, \citenamefont
  {Millis}, \citenamefont {Werner}, \citenamefont {Sood},\ and\ \citenamefont
  {Rao}}]{BretscherSA2021}%
  \BibitemOpen
  \bibfield  {author} {\bibinfo {author} {\bibfnamefont {H.~M.}\ \bibnamefont
  {Bretscher}}, \bibinfo {author} {\bibfnamefont {P.}~\bibnamefont {Andrich}},
  \bibinfo {author} {\bibfnamefont {Y.}~\bibnamefont {Murakami}}, \bibinfo
  {author} {\bibfnamefont {D.}~\bibnamefont {Golez}}, \bibinfo {author}
  {\bibfnamefont {B.}~\bibnamefont {Remez}}, \bibinfo {author} {\bibfnamefont
  {P.}~\bibnamefont {Telang}}, \bibinfo {author} {\bibfnamefont
  {A.}~\bibnamefont {Singh}}, \bibinfo {author} {\bibfnamefont
  {L.}~\bibnamefont {Harnagea}}, \bibinfo {author} {\bibfnamefont {N.~R.}\
  \bibnamefont {Cooper}}, \bibinfo {author} {\bibfnamefont {A.~J.}\
  \bibnamefont {Millis}}, \bibinfo {author} {\bibfnamefont {P.}~\bibnamefont
  {Werner}}, \bibinfo {author} {\bibfnamefont {A.~K.}\ \bibnamefont {Sood}},\
  and\ \bibinfo {author} {\bibfnamefont {A.}~\bibnamefont {Rao}},\ }\bibfield
  {title} {\bibinfo {title} {\rm{Imaging the coherent propagation of collective
  modes in the excitonic insulator Ta$_{2}$NiSe$_{5}$ at room temperature}},\
  }\href {https://doi.org/10.1126/sciadv.abd6147} {\bibfield  {journal}
  {\bibinfo  {journal} {Sci. Adv.}\ }\textbf {\bibinfo {volume} {7}},\ \bibinfo
  {pages} {eabd6147} (\bibinfo {year} {2021})}\BibitemShut {NoStop}%
\bibitem [{\citenamefont {Yuan}\ \emph {et~al.}(2018)\citenamefont {Yuan},
  \citenamefont {Zhu}, \citenamefont {Su}, \citenamefont {Yao}, \citenamefont
  {Xing}, \citenamefont {Chen}, \citenamefont {Ma}, \citenamefont {Lin},
  \citenamefont {Shi}, \citenamefont {Shindou}, \citenamefont {Xie},\ and\
  \citenamefont {Han}}]{YuanSA2018}%
  \BibitemOpen
  \bibfield  {author} {\bibinfo {author} {\bibfnamefont {W.}~\bibnamefont
  {Yuan}}, \bibinfo {author} {\bibfnamefont {Q.}~\bibnamefont {Zhu}}, \bibinfo
  {author} {\bibfnamefont {T.}~\bibnamefont {Su}}, \bibinfo {author}
  {\bibfnamefont {Y.}~\bibnamefont {Yao}}, \bibinfo {author} {\bibfnamefont
  {W.}~\bibnamefont {Xing}}, \bibinfo {author} {\bibfnamefont {Y.}~\bibnamefont
  {Chen}}, \bibinfo {author} {\bibfnamefont {Y.}~\bibnamefont {Ma}}, \bibinfo
  {author} {\bibfnamefont {X.}~\bibnamefont {Lin}}, \bibinfo {author}
  {\bibfnamefont {J.}~\bibnamefont {Shi}}, \bibinfo {author} {\bibfnamefont
  {R.}~\bibnamefont {Shindou}}, \bibinfo {author} {\bibfnamefont {X.~C.}\
  \bibnamefont {Xie}},\ and\ \bibinfo {author} {\bibfnamefont {W.}~\bibnamefont
  {Han}},\ }\bibfield  {title} {\bibinfo {title} {\rm{Experimental signatures
  of spin superfluid ground state in canted antiferromagnet Cr$_{2}$O$_{3}$ via
  nonlocal spin transport}},\ }\href {https://doi.org/10.1126/sciadv.aat1098}
  {\bibfield  {journal} {\bibinfo  {journal} {Sci. Adv.}\ }\textbf {\bibinfo
  {volume} {4}},\ \bibinfo {pages} {eaat1098} (\bibinfo {year}
  {2018})}\BibitemShut {NoStop}%
\bibitem [{\citenamefont {Jiang}\ \emph {et~al.}(2020)\citenamefont {Jiang},
  \citenamefont {Lou}, \citenamefont {Liu}, \citenamefont {Li}, \citenamefont
  {Song}, \citenamefont {Chang}, \citenamefont {Duan},\ and\ \citenamefont
  {Zhang}}]{JiangPRL2020}%
  \BibitemOpen
  \bibfield  {author} {\bibinfo {author} {\bibfnamefont {Z.}~\bibnamefont
  {Jiang}}, \bibinfo {author} {\bibfnamefont {W.}~\bibnamefont {Lou}}, \bibinfo
  {author} {\bibfnamefont {Y.}~\bibnamefont {Liu}}, \bibinfo {author}
  {\bibfnamefont {Y.}~\bibnamefont {Li}}, \bibinfo {author} {\bibfnamefont
  {H.}~\bibnamefont {Song}}, \bibinfo {author} {\bibfnamefont {K.}~\bibnamefont
  {Chang}}, \bibinfo {author} {\bibfnamefont {W.}~\bibnamefont {Duan}},\ and\
  \bibinfo {author} {\bibfnamefont {S.}~\bibnamefont {Zhang}},\ }\bibfield
  {title} {\bibinfo {title} {\rm{Spin-Triplet Excitonic Insulator: The Case of
  Semihydrogenated Graphene}},\ }\href
  {https://doi.org/10.1103/PhysRevLett.124.166401} {\bibfield  {journal}
  {\bibinfo  {journal} {Phys. Rev. Lett.}\ }\textbf {\bibinfo {volume} {124}},\
  \bibinfo {pages} {166401} (\bibinfo {year} {2020})}\BibitemShut {NoStop}%
\bibitem [{\citenamefont {Nasu}\ and\ \citenamefont
  {Naka}(2021)}]{NasuPRB2021}%
  \BibitemOpen
  \bibfield  {author} {\bibinfo {author} {\bibfnamefont {J.}~\bibnamefont
  {Nasu}}\ and\ \bibinfo {author} {\bibfnamefont {M.}~\bibnamefont {Naka}},\
  }\bibfield  {title} {\bibinfo {title} {Spin seebeck effect in nonmagnetic
  excitonic insulators},\ }\href {https://doi.org/10.1103/PhysRevB.103.L121104}
  {\bibfield  {journal} {\bibinfo  {journal} {Phys. Rev. B}\ }\textbf {\bibinfo
  {volume} {103}},\ \bibinfo {pages} {L121104} (\bibinfo {year}
  {2021})}\BibitemShut {NoStop}%
\bibitem [{\citenamefont {Manago}\ \emph {et~al.}(2019)\citenamefont {Manago},
  \citenamefont {Kitagawa}, \citenamefont {Ishida}, \citenamefont {Deguchi},
  \citenamefont {Sato},\ and\ \citenamefont {Yamamura}}]{Manago2019}%
  \BibitemOpen
  \bibfield  {author} {\bibinfo {author} {\bibfnamefont {M.}~\bibnamefont
  {Manago}}, \bibinfo {author} {\bibfnamefont {S.}~\bibnamefont {Kitagawa}},
  \bibinfo {author} {\bibfnamefont {K.}~\bibnamefont {Ishida}}, \bibinfo
  {author} {\bibfnamefont {K.}~\bibnamefont {Deguchi}}, \bibinfo {author}
  {\bibfnamefont {N.~K.}\ \bibnamefont {Sato}},\ and\ \bibinfo {author}
  {\bibfnamefont {T.}~\bibnamefont {Yamamura}},\ }\bibfield  {title} {\bibinfo
  {title} {\rm{Spin-triplet superconductivity in the paramagnetic UCoGe under
  pressure studied by $^{59}$Co NMR}},\ }\href
  {https://doi.org/10.1103/PhysRevB.100.035203} {\bibfield  {journal} {\bibinfo
   {journal} {Phys. Rev. B}\ }\textbf {\bibinfo {volume} {100}},\ \bibinfo
  {pages} {035203} (\bibinfo {year} {2019})}\BibitemShut {NoStop}%
\bibitem [{\citenamefont {Ran}\ \emph {et~al.}(2019)\citenamefont {Ran},
  \citenamefont {Eckberg}, \citenamefont {Ding}, \citenamefont {Furukawa},
  \citenamefont {Metz}, \citenamefont {Saha}, \citenamefont {Liu},
  \citenamefont {Zic}, \citenamefont {Kim}, \citenamefont {Paglione},\ and\
  \citenamefont {Butch}}]{Ran2019}%
  \BibitemOpen
  \bibfield  {author} {\bibinfo {author} {\bibfnamefont {S.}~\bibnamefont
  {Ran}}, \bibinfo {author} {\bibfnamefont {C.}~\bibnamefont {Eckberg}},
  \bibinfo {author} {\bibfnamefont {Q.~P.}\ \bibnamefont {Ding}}, \bibinfo
  {author} {\bibfnamefont {Y.}~\bibnamefont {Furukawa}}, \bibinfo {author}
  {\bibfnamefont {T.}~\bibnamefont {Metz}}, \bibinfo {author} {\bibfnamefont
  {S.~R.}\ \bibnamefont {Saha}}, \bibinfo {author} {\bibfnamefont {I.~L.}\
  \bibnamefont {Liu}}, \bibinfo {author} {\bibfnamefont {M.}~\bibnamefont
  {Zic}}, \bibinfo {author} {\bibfnamefont {H.}~\bibnamefont {Kim}}, \bibinfo
  {author} {\bibfnamefont {J.}~\bibnamefont {Paglione}},\ and\ \bibinfo
  {author} {\bibfnamefont {N.~P.}\ \bibnamefont {Butch}},\ }\bibfield  {title}
  {\bibinfo {title} {Nearly ferromagnetic spin-triplet superconductivity},\
  }\href {https://doi.org/10.1126/science.aav8645} {\bibfield  {journal}
  {\bibinfo  {journal} {Science}\ }\textbf {\bibinfo {volume} {365}},\ \bibinfo
  {pages} {684} (\bibinfo {year} {2019})}\BibitemShut {NoStop}%
\bibitem [{\citenamefont {Halperin}(2019)}]{HalperinARCMP2019}%
  \BibitemOpen
  \bibfield  {author} {\bibinfo {author} {\bibfnamefont {W.~P.}\ \bibnamefont
  {Halperin}},\ }\bibfield  {title} {\bibinfo {title} {\rm{Superfluid $^{3}$He
  in Aerogel}},\ }\href
  {https://doi.org/10.1146/annurev-conmatphys-031218-013134} {\bibfield
  {journal} {\bibinfo  {journal} {Annu. Rev. Condens. Matter Phys.}\ }\textbf
  {\bibinfo {volume} {10}},\ \bibinfo {pages} {155} (\bibinfo {year}
  {2019})}\BibitemShut {NoStop}%
\bibitem [{\citenamefont {Kaneko}\ \emph {et~al.}(2015)\citenamefont {Kaneko},
  \citenamefont {Zenker}, \citenamefont {Fehske},\ and\ \citenamefont
  {Ohta}}]{KanekoPRB2015}%
  \BibitemOpen
  \bibfield  {author} {\bibinfo {author} {\bibfnamefont {T.}~\bibnamefont
  {Kaneko}}, \bibinfo {author} {\bibfnamefont {B.}~\bibnamefont {Zenker}},
  \bibinfo {author} {\bibfnamefont {H.}~\bibnamefont {Fehske}},\ and\ \bibinfo
  {author} {\bibfnamefont {Y.}~\bibnamefont {Ohta}},\ }\bibfield  {title}
  {\bibinfo {title} {\rm{Competition between excitonic charge and spin density
  waves: Influence of electron-phonon and Hund's rule couplings}},\ }\href
  {https://doi.org/10.1103/PhysRevB.92.115106} {\bibfield  {journal} {\bibinfo
  {journal} {Phys. Rev. B}\ }\textbf {\bibinfo {volume} {92}},\ \bibinfo
  {pages} {115106} (\bibinfo {year} {2015})}\BibitemShut {NoStop}%
\bibitem [{\citenamefont {Kune\v{s}}(2015)}]{KunesJPCM2015}%
  \BibitemOpen
  \bibfield  {author} {\bibinfo {author} {\bibfnamefont {J.}~\bibnamefont
  {Kune\v{s}}},\ }\bibfield  {title} {\bibinfo {title} {Excitonic condensation
  in systems of strongly correlated electrons},\ }\href
  {https://doi.org/10.1088/0953-8984/27/33/333201} {\bibfield  {journal}
  {\bibinfo  {journal} {J. Phys. Cond. Mat.}\ }\textbf {\bibinfo {volume}
  {27}},\ \bibinfo {pages} {333201} (\bibinfo {year} {2015})}\BibitemShut
  {NoStop}%
\bibitem [{\citenamefont {Nasu}\ \emph {et~al.}(2016)\citenamefont {Nasu},
  \citenamefont {Watanabe}, \citenamefont {Naka},\ and\ \citenamefont
  {Ishihara}}]{NasuPRB2016}%
  \BibitemOpen
  \bibfield  {author} {\bibinfo {author} {\bibfnamefont {J.}~\bibnamefont
  {Nasu}}, \bibinfo {author} {\bibfnamefont {T.}~\bibnamefont {Watanabe}},
  \bibinfo {author} {\bibfnamefont {M.}~\bibnamefont {Naka}},\ and\ \bibinfo
  {author} {\bibfnamefont {S.}~\bibnamefont {Ishihara}},\ }\bibfield  {title}
  {\bibinfo {title} {Phase diagram and collective excitations in an excitonic
  insulator from an orbital physics viewpoint},\ }\href
  {https://doi.org/10.1103/PhysRevB.93.205136} {\bibfield  {journal} {\bibinfo
  {journal} {Phys. Rev. B}\ }\textbf {\bibinfo {volume} {93}},\ \bibinfo
  {pages} {205136} (\bibinfo {year} {2016})}\BibitemShut {NoStop}%
\bibitem [{\citenamefont {Ikeda}\ \emph {et~al.}(2020)\citenamefont {Ikeda},
  \citenamefont {Matsuda},\ and\ \citenamefont {Sato}}]{IkedaPRL2020}%
  \BibitemOpen
  \bibfield  {author} {\bibinfo {author} {\bibfnamefont {A.}~\bibnamefont
  {Ikeda}}, \bibinfo {author} {\bibfnamefont {Y.~H.}\ \bibnamefont {Matsuda}},\
  and\ \bibinfo {author} {\bibfnamefont {K.}~\bibnamefont {Sato}},\ }\bibfield
  {title} {\bibinfo {title} {Two spin-state crystallizations in
  \rm{LaCoO$_{3}$}},\ }\href {https://doi.org/10.1103/PhysRevLett.125.177202}
  {\bibfield  {journal} {\bibinfo  {journal} {Phys. Rev. Lett.}\ }\textbf
  {\bibinfo {volume} {125}},\ \bibinfo {pages} {177202} (\bibinfo {year}
  {2020})}\BibitemShut {NoStop}%
\bibitem [{\citenamefont {Ikeda}\ \emph {et~al.}(2016)\citenamefont {Ikeda},
  \citenamefont {Nomura}, \citenamefont {Matsuda}, \citenamefont {Matsuo},
  \citenamefont {Kindo},\ and\ \citenamefont {Sato}}]{IkedaLCO2016}%
  \BibitemOpen
  \bibfield  {author} {\bibinfo {author} {\bibfnamefont {A.}~\bibnamefont
  {Ikeda}}, \bibinfo {author} {\bibfnamefont {T.}~\bibnamefont {Nomura}},
  \bibinfo {author} {\bibfnamefont {Y.~H.}\ \bibnamefont {Matsuda}}, \bibinfo
  {author} {\bibfnamefont {A.}~\bibnamefont {Matsuo}}, \bibinfo {author}
  {\bibfnamefont {K.}~\bibnamefont {Kindo}},\ and\ \bibinfo {author}
  {\bibfnamefont {K.}~\bibnamefont {Sato}},\ }\bibfield  {title} {\bibinfo
  {title} {Spin state ordering of strongly correlating $\rm{LaCoO}_{3}$ induced
  at ultrahigh magnetic fields},\ }\href
  {https://doi.org/10.1103/PhysRevB.93.220401} {\bibfield  {journal} {\bibinfo
  {journal} {Phys. Rev. B}\ }\textbf {\bibinfo {volume} {93}},\ \bibinfo
  {pages} {220401(R)} (\bibinfo {year} {2016})}\BibitemShut {NoStop}%
\bibitem [{\citenamefont {Hejtm\'{a}nek}\ \emph {et~al.}(2013)\citenamefont
  {Hejtm\'{a}nek}, \citenamefont {Jir\'{a}k}, \citenamefont {Kaman},
  \citenamefont {Kn\'{i}\v{z}ek}, \citenamefont {\v{S}antav\'{a}},
  \citenamefont {Nitta}, \citenamefont {Naito},\ and\ \citenamefont
  {Fujishiro}}]{HejtmanekEPJB2013}%
  \BibitemOpen
  \bibfield  {author} {\bibinfo {author} {\bibfnamefont {J.}~\bibnamefont
  {Hejtm\'{a}nek}}, \bibinfo {author} {\bibfnamefont {Z.}~\bibnamefont
  {Jir\'{a}k}}, \bibinfo {author} {\bibfnamefont {O.}~\bibnamefont {Kaman}},
  \bibinfo {author} {\bibfnamefont {K.}~\bibnamefont {Kn\'{i}\v{z}ek}},
  \bibinfo {author} {\bibfnamefont {E.}~\bibnamefont {\v{S}antav\'{a}}},
  \bibinfo {author} {\bibfnamefont {K.}~\bibnamefont {Nitta}}, \bibinfo
  {author} {\bibfnamefont {T.}~\bibnamefont {Naito}},\ and\ \bibinfo {author}
  {\bibfnamefont {H.}~\bibnamefont {Fujishiro}},\ }\bibfield  {title} {\bibinfo
  {title} {\rm{Phase transition in Pr$_{0.5}$Ca$_{0.5}$CoO$_{3}$ and related
  cobaltites}},\ }\href {https://doi.org/10.1140/epjb/e2013-30653-y} {\bibfield
   {journal} {\bibinfo  {journal} {Eur. Phys. J. B}\ }\textbf {\bibinfo
  {volume} {86}},\ \bibinfo {pages} {305} (\bibinfo {year} {2013})}\BibitemShut
  {NoStop}%
\bibitem [{\citenamefont {Kune\v{s}}\ and\ \citenamefont
  {Augustinsk\'{y}}(2014)}]{KunesPRB2014}%
  \BibitemOpen
  \bibfield  {author} {\bibinfo {author} {\bibfnamefont {J.}~\bibnamefont
  {Kune\v{s}}}\ and\ \bibinfo {author} {\bibfnamefont {P.}~\bibnamefont
  {Augustinsk\'{y}}},\ }\bibfield  {title} {\bibinfo {title} {Excitonic
  instability at the spin-state transition in the two-band hubbard model},\
  }\href {https://doi.org/10.1103/PhysRevB.89.115134} {\bibfield  {journal}
  {\bibinfo  {journal} {Phys. Rev. B}\ }\textbf {\bibinfo {volume} {89}},\
  \bibinfo {pages} {115134} (\bibinfo {year} {2014})}\BibitemShut {NoStop}%
\bibitem [{\citenamefont {Sotnikov}\ and\ \citenamefont
  {Kune\v{s}}(2016)}]{SotnikovSR2016}%
  \BibitemOpen
  \bibfield  {author} {\bibinfo {author} {\bibfnamefont {A.}~\bibnamefont
  {Sotnikov}}\ and\ \bibinfo {author} {\bibfnamefont {J.}~\bibnamefont
  {Kune\v{s}}},\ }\bibfield  {title} {\bibinfo {title} {\rm{Field-induced
  exciton condensation in LaCoO$_{3}$}},\ }\href
  {https://doi.org/10.1038/srep30510} {\bibfield  {journal} {\bibinfo
  {journal} {Sci. Rep.}\ }\textbf {\bibinfo {volume} {6}},\ \bibinfo {pages}
  {30510} (\bibinfo {year} {2016})}\BibitemShut {NoStop}%
\bibitem [{\citenamefont {Tatsuno}\ \emph {et~al.}(2016)\citenamefont
  {Tatsuno}, \citenamefont {Mizoguchi}, \citenamefont {Nasu}, \citenamefont
  {Naka},\ and\ \citenamefont {Ishihara}}]{TatsunoJPSJ2016}%
  \BibitemOpen
  \bibfield  {author} {\bibinfo {author} {\bibfnamefont {T.}~\bibnamefont
  {Tatsuno}}, \bibinfo {author} {\bibfnamefont {E.}~\bibnamefont {Mizoguchi}},
  \bibinfo {author} {\bibfnamefont {J.}~\bibnamefont {Nasu}}, \bibinfo {author}
  {\bibfnamefont {M.}~\bibnamefont {Naka}},\ and\ \bibinfo {author}
  {\bibfnamefont {S.}~\bibnamefont {Ishihara}},\ }\bibfield  {title} {\bibinfo
  {title} {Magnetic field effects in a correlated electron system with
  spin-state degree of freedom ― implications for an excitonic insulator ―},\
  }\href {https://doi.org/10.7566/JPSJ.85.083706} {\bibfield  {journal}
  {\bibinfo  {journal} {J. Phys. Soc. Jpn.}\ }\textbf {\bibinfo {volume}
  {85}},\ \bibinfo {pages} {083706} (\bibinfo {year} {2016})}\BibitemShut
  {NoStop}%
\bibitem [{\citenamefont {Nakamura}\ \emph {et~al.}(2018)\citenamefont
  {Nakamura}, \citenamefont {Ikeda}, \citenamefont {Sawabe}, \citenamefont
  {Matsuda},\ and\ \citenamefont {Takeyama}}]{NakamuraRSI2018}%
  \BibitemOpen
  \bibfield  {author} {\bibinfo {author} {\bibfnamefont {D.}~\bibnamefont
  {Nakamura}}, \bibinfo {author} {\bibfnamefont {A.}~\bibnamefont {Ikeda}},
  \bibinfo {author} {\bibfnamefont {H.}~\bibnamefont {Sawabe}}, \bibinfo
  {author} {\bibfnamefont {Y.~H.}\ \bibnamefont {Matsuda}},\ and\ \bibinfo
  {author} {\bibfnamefont {S.}~\bibnamefont {Takeyama}},\ }\bibfield  {title}
  {\bibinfo {title} {Record indoor magnetic field of \rm{1200 T} generated by
  electromagnetic flux-compression},\ }\href
  {https://doi.org/10.1063/1.5044557} {\bibfield  {journal} {\bibinfo
  {journal} {Rev. Sci. Instrum.}\ }\textbf {\bibinfo {volume} {89}},\ \bibinfo
  {pages} {095106} (\bibinfo {year} {2018})}\BibitemShut {NoStop}%
\bibitem [{\citenamefont {Ikeda}\ \emph {et~al.}(2017)\citenamefont {Ikeda},
  \citenamefont {Nomura}, \citenamefont {Matsuda}, \citenamefont {Tani},
  \citenamefont {Kobayashi}, \citenamefont {Watanabe},\ and\ \citenamefont
  {Sato}}]{AIkedaFBG2017}%
  \BibitemOpen
  \bibfield  {author} {\bibinfo {author} {\bibfnamefont {A.}~\bibnamefont
  {Ikeda}}, \bibinfo {author} {\bibfnamefont {T.}~\bibnamefont {Nomura}},
  \bibinfo {author} {\bibfnamefont {Y.~H.}\ \bibnamefont {Matsuda}}, \bibinfo
  {author} {\bibfnamefont {S.}~\bibnamefont {Tani}}, \bibinfo {author}
  {\bibfnamefont {Y.}~\bibnamefont {Kobayashi}}, \bibinfo {author}
  {\bibfnamefont {H.}~\bibnamefont {Watanabe}},\ and\ \bibinfo {author}
  {\bibfnamefont {K.}~\bibnamefont {Sato}},\ }\bibfield  {title} {\bibinfo
  {title} {High-speed \rm{100 MHz} strain monitor using fiber \rm{Bragg}
  grating and optical filter for magnetostriction measurements under ultrahigh
  magnetic fields},\ }\href {https://doi.org/10.1063/1.4999452} {\bibfield
  {journal} {\bibinfo  {journal} {Rev. Sci. Instrum.}\ }\textbf {\bibinfo
  {volume} {88}},\ \bibinfo {pages} {083906} (\bibinfo {year}
  {2017})}\BibitemShut {NoStop}%
\bibitem [{\citenamefont {Sotnikov}\ and\ \citenamefont
  {Kune\v{s}}(2017)}]{SotnikovPRB2017}%
  \BibitemOpen
  \bibfield  {author} {\bibinfo {author} {\bibfnamefont {A.}~\bibnamefont
  {Sotnikov}}\ and\ \bibinfo {author} {\bibfnamefont {J.}~\bibnamefont
  {Kune\v{s}}},\ }\bibfield  {title} {\bibinfo {title} {\rm{Competing phases in
  a model of Pr-based cobaltites}},\ }\href
  {https://doi.org/10.1103/PhysRevB.96.245102} {\bibfield  {journal} {\bibinfo
  {journal} {Phys. Rev. B}\ }\textbf {\bibinfo {volume} {96}},\ \bibinfo
  {pages} {245102} (\bibinfo {year} {2017})}\BibitemShut {NoStop}%
\bibitem [{\citenamefont {Rice}(2002)}]{RiceScience2002}%
  \BibitemOpen
  \bibfield  {author} {\bibinfo {author} {\bibfnamefont {T.~M.}\ \bibnamefont
  {Rice}},\ }\bibfield  {title} {\bibinfo {title} {\rm{Quantum mechanics. To
  condense or not to condense}},\ }\href
  {https://doi.org/10.1126/science.1078819} {\bibfield  {journal} {\bibinfo
  {journal} {Science}\ }\textbf {\bibinfo {volume} {298}},\ \bibinfo {pages}
  {760} (\bibinfo {year} {2002})}\BibitemShut {NoStop}%
\bibitem [{\citenamefont {Zapf}\ \emph {et~al.}(2014)\citenamefont {Zapf},
  \citenamefont {Jaime},\ and\ \citenamefont {Batista}}]{ZapfRMP2014}%
  \BibitemOpen
  \bibfield  {author} {\bibinfo {author} {\bibfnamefont {V.}~\bibnamefont
  {Zapf}}, \bibinfo {author} {\bibfnamefont {M.}~\bibnamefont {Jaime}},\ and\
  \bibinfo {author} {\bibfnamefont {C.~U.}\ \bibnamefont {Batista}},\
  }\bibfield  {title} {\bibinfo {title} {\rm{Bose-Einstein condensation in
  quantum magnets}},\ }\href {https://doi.org/10.1103/RevModPhys.86.563}
  {\bibfield  {journal} {\bibinfo  {journal} {Rev. Mod. Phys.}\ }\textbf
  {\bibinfo {volume} {86}},\ \bibinfo {pages} {563} (\bibinfo {year}
  {2014})}\BibitemShut {NoStop}%
\bibitem [{\citenamefont {Sato}\ \emph {et~al.}(2009)\citenamefont {Sato},
  \citenamefont {Matsuo}, \citenamefont {Kindo}, \citenamefont {Kobayashi},\
  and\ \citenamefont {Asai}}]{SatoJPSJ2009}%
  \BibitemOpen
  \bibfield  {author} {\bibinfo {author} {\bibfnamefont {K.}~\bibnamefont
  {Sato}}, \bibinfo {author} {\bibfnamefont {A.}~\bibnamefont {Matsuo}},
  \bibinfo {author} {\bibfnamefont {K.}~\bibnamefont {Kindo}}, \bibinfo
  {author} {\bibfnamefont {Y.}~\bibnamefont {Kobayashi}},\ and\ \bibinfo
  {author} {\bibfnamefont {K.}~\bibnamefont {Asai}},\ }\bibfield  {title}
  {\bibinfo {title} {Field induced spin-state transition in $\rm{LaCoO}_{3}$},\
  }\href {https://doi.org/10.1143/JPSJ.80.104702} {\bibfield  {journal}
  {\bibinfo  {journal} {J. Phys. Soc. Jpn.}\ }\textbf {\bibinfo {volume}
  {78}},\ \bibinfo {pages} {093702} (\bibinfo {year} {2009})}\BibitemShut
  {NoStop}%
\bibitem [{\citenamefont {Altarawneh}\ \emph {et~al.}(2012)\citenamefont
  {Altarawneh}, \citenamefont {Chern}, \citenamefont {Harrison}, \citenamefont
  {Batista}, \citenamefont {Uchida}, \citenamefont {Jaime}, \citenamefont
  {Rickel}, \citenamefont {Crooker}, \citenamefont {Mielke}, \citenamefont
  {Betts}, \citenamefont {Mitchell},\ and\ \citenamefont {Hoch}}]{MoazPRL2012}%
  \BibitemOpen
  \bibfield  {author} {\bibinfo {author} {\bibfnamefont {M.~M.}\ \bibnamefont
  {Altarawneh}}, \bibinfo {author} {\bibfnamefont {G.~W.}\ \bibnamefont
  {Chern}}, \bibinfo {author} {\bibfnamefont {N.}~\bibnamefont {Harrison}},
  \bibinfo {author} {\bibfnamefont {C.~D.}\ \bibnamefont {Batista}}, \bibinfo
  {author} {\bibfnamefont {A.}~\bibnamefont {Uchida}}, \bibinfo {author}
  {\bibfnamefont {M.}~\bibnamefont {Jaime}}, \bibinfo {author} {\bibfnamefont
  {D.~G.}\ \bibnamefont {Rickel}}, \bibinfo {author} {\bibfnamefont {S.~A.}\
  \bibnamefont {Crooker}}, \bibinfo {author} {\bibfnamefont {C.~H.}\
  \bibnamefont {Mielke}}, \bibinfo {author} {\bibfnamefont {J.~B.}\
  \bibnamefont {Betts}}, \bibinfo {author} {\bibfnamefont {J.~F.}\ \bibnamefont
  {Mitchell}},\ and\ \bibinfo {author} {\bibfnamefont {M.~J.~R.}\ \bibnamefont
  {Hoch}},\ }\bibfield  {title} {\bibinfo {title} {Cascade of magnetic field
  induced spin transitions in $\rm{LaCoO}_{3}$},\ }\href
  {https://doi.org/10.1103/PhysRevLett.109.037201} {\bibfield  {journal}
  {\bibinfo  {journal} {Phys. Rev. Lett.}\ }\textbf {\bibinfo {volume} {109}},\
  \bibinfo {pages} {037201} (\bibinfo {year} {2012})}\BibitemShut {NoStop}%
\bibitem [{\citenamefont {Rotter}\ \emph {et~al.}(2014)\citenamefont {Rotter},
  \citenamefont {Wang}, \citenamefont {Boothroyd}, \citenamefont {Prabhakaran},
  \citenamefont {Tanaka},\ and\ \citenamefont {Doerr}}]{RotterSR2014}%
  \BibitemOpen
  \bibfield  {author} {\bibinfo {author} {\bibfnamefont {M.}~\bibnamefont
  {Rotter}}, \bibinfo {author} {\bibfnamefont {Z.~S.}\ \bibnamefont {Wang}},
  \bibinfo {author} {\bibfnamefont {A.~T.}\ \bibnamefont {Boothroyd}}, \bibinfo
  {author} {\bibfnamefont {D.}~\bibnamefont {Prabhakaran}}, \bibinfo {author}
  {\bibfnamefont {A.}~\bibnamefont {Tanaka}},\ and\ \bibinfo {author}
  {\bibfnamefont {M.}~\bibnamefont {Doerr}},\ }\bibfield  {title} {\bibinfo
  {title} {\rm{Mechanism of spin crossover in LaCoO$_{3}$ resolved by shape
  magnetostriction in pulsed magnetic fields}},\ }\href
  {https://doi.org/10.1038/srep07003} {\bibfield  {journal} {\bibinfo
  {journal} {Sci. Rep.}\ }\textbf {\bibinfo {volume} {4}},\ \bibinfo {pages}
  {7003} (\bibinfo {year} {2014})}\BibitemShut {NoStop}%
\bibitem [{\citenamefont {Platonov}\ \emph {et~al.}(2012)\citenamefont
  {Platonov}, \citenamefont {Kudasov}, \citenamefont {Monakhov},\ and\
  \citenamefont {Tatsenko}}]{Platonov2012}%
  \BibitemOpen
  \bibfield  {author} {\bibinfo {author} {\bibfnamefont {V.~V.}\ \bibnamefont
  {Platonov}}, \bibinfo {author} {\bibfnamefont {Y.~B.}\ \bibnamefont
  {Kudasov}}, \bibinfo {author} {\bibfnamefont {M.~P.}\ \bibnamefont
  {Monakhov}},\ and\ \bibinfo {author} {\bibfnamefont {O.~M.}\ \bibnamefont
  {Tatsenko}},\ }\bibfield  {title} {\bibinfo {title} {\rm{Magnetically induced
  phase transitions in LaCoO$_{3}$ in fields of up to 500 T}},\ }\href
  {https://doi.org/10.1134/S1063783412020266} {\bibfield  {journal} {\bibinfo
  {journal} {Phys. of the Solid State}\ }\textbf {\bibinfo {volume} {54}},\
  \bibinfo {pages} {279} (\bibinfo {year} {2012})}\BibitemShut {NoStop}%
\bibitem [{\citenamefont {Sch\"{o}nemann}\ \emph {et~al.}(2021)\citenamefont
  {Sch\"{o}nemann}, \citenamefont {Rodriguez}, \citenamefont {Rickel},
  \citenamefont {Balakirev}, \citenamefont {McDonald}, \citenamefont {Evans},
  \citenamefont {Maiorov}, \citenamefont {Paillard}, \citenamefont {Bellaiche},
  \citenamefont {Stier}, \citenamefont {Salamon}, \citenamefont {Gofryk},\ and\
  \citenamefont {Jaime}}]{RicoPNAS}%
  \BibitemOpen
  \bibfield  {author} {\bibinfo {author} {\bibfnamefont {R.}~\bibnamefont
  {Sch\"{o}nemann}}, \bibinfo {author} {\bibfnamefont {G.}~\bibnamefont
  {Rodriguez}}, \bibinfo {author} {\bibfnamefont {D.}~\bibnamefont {Rickel}},
  \bibinfo {author} {\bibfnamefont {F.}~\bibnamefont {Balakirev}}, \bibinfo
  {author} {\bibfnamefont {R.~D.}\ \bibnamefont {McDonald}}, \bibinfo {author}
  {\bibfnamefont {J.~A.}\ \bibnamefont {Evans}}, \bibinfo {author}
  {\bibfnamefont {B.}~\bibnamefont {Maiorov}}, \bibinfo {author} {\bibfnamefont
  {C.}~\bibnamefont {Paillard}}, \bibinfo {author} {\bibfnamefont
  {L.}~\bibnamefont {Bellaiche}}, \bibinfo {author} {\bibfnamefont {A.~V.}\
  \bibnamefont {Stier}}, \bibinfo {author} {\bibfnamefont {M.~B.}\ \bibnamefont
  {Salamon}}, \bibinfo {author} {\bibfnamefont {K.}~\bibnamefont {Gofryk}},\
  and\ \bibinfo {author} {\bibfnamefont {M.}~\bibnamefont {Jaime}},\ }\bibfield
   {title} {\bibinfo {title} {\rm{Magnetoelastic standing waves induced in
  UO$_{2}$ by microsecond magnetic field pulses}},\ }\href
  {https://doi.org/10.1073/pnas.2110555118} {\bibfield  {journal} {\bibinfo
  {journal} {Proc. Natl. Acad. Sci. U. S. A.}\ }\textbf {\bibinfo {volume}
  {118}},\ \bibinfo {pages} {e2110555118} (\bibinfo {year} {2021})}\BibitemShut
  {NoStop}%
\bibitem [{\citenamefont {Ky\^{o}men}\ \emph {et~al.}(2003)\citenamefont
  {Ky\^{o}men}, \citenamefont {Asaka},\ and\ \citenamefont
  {Itoh}}]{KyomenPRB2003}%
  \BibitemOpen
  \bibfield  {author} {\bibinfo {author} {\bibfnamefont {T.}~\bibnamefont
  {Ky\^{o}men}}, \bibinfo {author} {\bibfnamefont {Y.}~\bibnamefont {Asaka}},\
  and\ \bibinfo {author} {\bibfnamefont {M.}~\bibnamefont {Itoh}},\ }\bibfield
  {title} {\bibinfo {title} {Negative cooperative effect on the spin-state
  excitation in $\rm{LaCoO}_{3}$},\ }\href
  {https://doi.org/10.1103/PhysRevB.67.144424} {\bibfield  {journal} {\bibinfo
  {journal} {Phys. Rev. B}\ }\textbf {\bibinfo {volume} {67}},\ \bibinfo
  {pages} {144424} (\bibinfo {year} {2003})}\BibitemShut {NoStop}%
\bibitem [{\citenamefont {Radaelli}\ and\ \citenamefont
  {Cheong}(2002)}]{Radaelli}%
  \BibitemOpen
  \bibfield  {author} {\bibinfo {author} {\bibfnamefont {P.~G.}\ \bibnamefont
  {Radaelli}}\ and\ \bibinfo {author} {\bibfnamefont {S.~W.}\ \bibnamefont
  {Cheong}},\ }\bibfield  {title} {\bibinfo {title} {Structural phenomena
  associated with the spin-state transition in $\rm{LaCoO}_{3}$},\ }\href
  {https://doi.org/10.1103/PhysRevB.66.094408} {\bibfield  {journal} {\bibinfo
  {journal} {Phys. Rev. B}\ }\textbf {\bibinfo {volume} {66}},\ \bibinfo
  {pages} {094408} (\bibinfo {year} {2002})}\BibitemShut {NoStop}%
\bibitem [{\citenamefont {Noguchi}\ \emph {et~al.}(2002)\citenamefont
  {Noguchi}, \citenamefont {Kawamata}, \citenamefont {Okuda}, \citenamefont
  {Nojiri},\ and\ \citenamefont {Motokawa}}]{NoguchiPRB2002}%
  \BibitemOpen
  \bibfield  {author} {\bibinfo {author} {\bibfnamefont {S.}~\bibnamefont
  {Noguchi}}, \bibinfo {author} {\bibfnamefont {S.}~\bibnamefont {Kawamata}},
  \bibinfo {author} {\bibfnamefont {K.}~\bibnamefont {Okuda}}, \bibinfo
  {author} {\bibfnamefont {H.}~\bibnamefont {Nojiri}},\ and\ \bibinfo {author}
  {\bibfnamefont {M.}~\bibnamefont {Motokawa}},\ }\bibfield  {title} {\bibinfo
  {title} {\rm{Evidence for the excited triplet of LaCoO}$_{3}$},\ }\href
  {https://doi.org/10.1103/PhysRevB.66.094404} {\bibfield  {journal} {\bibinfo
  {journal} {Phys. Rev. B}\ }\textbf {\bibinfo {volume} {66}},\ \bibinfo
  {pages} {094404} (\bibinfo {year} {2002})}\BibitemShut {NoStop}%
\bibitem [{\citenamefont {Okimoto}\ \emph {et~al.}(2021)\citenamefont
  {Okimoto}, \citenamefont {Saitoh}, \citenamefont {Kobayashi},\ and\
  \citenamefont {Ishihara}}]{SCObook}%
  \BibitemOpen
  \bibinfo {editor} {\bibfnamefont {Y.}~\bibnamefont {Okimoto}}, \bibinfo
  {editor} {\bibfnamefont {T.}~\bibnamefont {Saitoh}}, \bibinfo {editor}
  {\bibfnamefont {Y.}~\bibnamefont {Kobayashi}},\ and\ \bibinfo {editor}
  {\bibfnamefont {S.}~\bibnamefont {Ishihara}},\ eds.,\ \href
  {https://doi.org/10.1007/978-981-15-7929-5} {\emph {\bibinfo {title}
  {Spin-Crossover Cobaltite (Review and Outlook)}}},\ Springer Series in
  Materials Science\ (\bibinfo  {publisher} {Springer Singapore},\ \bibinfo
  {year} {2021})\BibitemShut {NoStop}%
\bibitem [{\citenamefont {Ikeda}\ \emph {et~al.}(2022)\citenamefont {Ikeda},
  \citenamefont {Matsuda}, \citenamefont {Zhou}, \citenamefont {Peng},
  \citenamefont {Ishii}, \citenamefont {Yajima}, \citenamefont {Kubota},
  \citenamefont {Inoue}, \citenamefont {Inubushi}, \citenamefont {Tono},\ and\
  \citenamefont {Yabashi}}]{IkedaAPL2022}%
  \BibitemOpen
  \bibfield  {author} {\bibinfo {author} {\bibfnamefont {A.}~\bibnamefont
  {Ikeda}}, \bibinfo {author} {\bibfnamefont {Y.~H.}\ \bibnamefont {Matsuda}},
  \bibinfo {author} {\bibfnamefont {X.}~\bibnamefont {Zhou}}, \bibinfo {author}
  {\bibfnamefont {S.}~\bibnamefont {Peng}}, \bibinfo {author} {\bibfnamefont
  {Y.}~\bibnamefont {Ishii}}, \bibinfo {author} {\bibfnamefont
  {T.}~\bibnamefont {Yajima}}, \bibinfo {author} {\bibfnamefont
  {Y.}~\bibnamefont {Kubota}}, \bibinfo {author} {\bibfnamefont
  {I.}~\bibnamefont {Inoue}}, \bibinfo {author} {\bibfnamefont
  {Y.}~\bibnamefont {Inubushi}}, \bibinfo {author} {\bibfnamefont
  {K.}~\bibnamefont {Tono}},\ and\ \bibinfo {author} {\bibfnamefont
  {M.}~\bibnamefont {Yabashi}},\ }\bibfield  {title} {\bibinfo {title}
  {\rm{Generating 77 T using a portable pulse magnet for single-shot quantum
  beam experiments}},\ }\href {https://doi.org/10.1063/5.0088134} {\bibfield
  {journal} {\bibinfo  {journal} {Appl. Phys. Lett.}\ }\textbf {\bibinfo
  {volume} {120}},\ \bibinfo {pages} {142403} (\bibinfo {year}
  {2022})}\BibitemShut {NoStop}%
\bibitem [{\citenamefont {Ikeda}(2023)}]{IkedaZENODO2023}%
  \BibitemOpen
  \bibfield  {author} {\bibinfo {author} {\bibfnamefont {A.}~\bibnamefont
  {Ikeda}},\ }\bibfield  {title} {\bibinfo {title} {\rm{Raw data for Signature
  of spin-triplet exciton condensations in LaCoO$_{3}$ at ultrahigh magnetic
  fields up to 600 T}},\ }\bibfield  {journal} {\bibinfo  {journal} {ZENODO}\
  }\href {https://doi.org/10.5281/zenodo.7627257} {10.5281/zenodo.7627257}
  (\bibinfo {year} {2023})\BibitemShut {NoStop}%
\end{thebibliography}%


\begin{thebibliography}{12}%
\makeatletter
\providecommand \@ifxundefined [1]{%
 \@ifx{#1\undefined}
}%
\providecommand \@ifnum [1]{%
 \ifnum #1\expandafter \@firstoftwo
 \else \expandafter \@secondoftwo
 \fi
}%
\providecommand \@ifx [1]{%
 \ifx #1\expandafter \@firstoftwo
 \else \expandafter \@secondoftwo
 \fi
}%
\providecommand \natexlab [1]{#1}%
\providecommand \enquote  [1]{``#1''}%
\providecommand \bibnamefont  [1]{#1}%
\providecommand \bibfnamefont [1]{#1}%
\providecommand \citenamefont [1]{#1}%
\providecommand \href@noop [0]{\@secondoftwo}%
\providecommand \href [0]{\begingroup \@sanitize@url \@href}%
\providecommand \@href[1]{\@@startlink{#1}\@@href}%
\providecommand \@@href[1]{\endgroup#1\@@endlink}%
\providecommand \@sanitize@url [0]{\catcode `\\12\catcode `\$12\catcode
  `\&12\catcode `\#12\catcode `\^12\catcode `\_12\catcode `\%12\relax}%
\providecommand \@@startlink[1]{}%
\providecommand \@@endlink[0]{}%
\providecommand \url  [0]{\begingroup\@sanitize@url \@url }%
\providecommand \@url [1]{\endgroup\@href {#1}{\urlprefix }}%
\providecommand \urlprefix  [0]{URL }%
\providecommand \Eprint [0]{\href }%
\providecommand \doibase [0]{http://dx.doi.org/}%
\providecommand \selectlanguage [0]{\@gobble}%
\providecommand \bibinfo  [0]{\@secondoftwo}%
\providecommand \bibfield  [0]{\@secondoftwo}%
\providecommand \translation [1]{[#1]}%
\providecommand \BibitemOpen [0]{}%
\providecommand \bibitemStop [0]{}%
\providecommand \bibitemNoStop [0]{.\EOS\space}%
\providecommand \EOS [0]{\spacefactor3000\relax}%
\providecommand \BibitemShut  [1]{\csname bibitem#1\endcsname}%
\let\auto@bib@innerbib\@empty
%</preamble>
\bibitem [{\citenamefont {Nakamura}\ \emph {et~al.}(2018)\citenamefont
  {Nakamura}, \citenamefont {Ikeda}, \citenamefont {Sawabe}, \citenamefont
  {Matsuda},\ and\ \citenamefont {Takeyama}}]{NakamuraRSI2018}%
  \BibitemOpen
  \bibfield  {author} {\bibinfo {author} {\bibfnamefont {D.}~\bibnamefont
  {Nakamura}}, \bibinfo {author} {\bibfnamefont {A.}~\bibnamefont {Ikeda}},
  \bibinfo {author} {\bibfnamefont {H.}~\bibnamefont {Sawabe}}, \bibinfo
  {author} {\bibfnamefont {Y.~H.}\ \bibnamefont {Matsuda}}, \ and\ \bibinfo
  {author} {\bibfnamefont {S.}~\bibnamefont {Takeyama}},\ }\href {\doibase
  10.1063/1.5044557} {\bibfield  {journal} {\bibinfo  {journal} {Rev. Sci.
  Instrum.}\ }\textbf {\bibinfo {volume} {89}},\ \bibinfo {pages} {095106}
  (\bibinfo {year} {2018})}\BibitemShut {NoStop}%
\bibitem [{\citenamefont {Ikeda}\ \emph {et~al.}(2017)\citenamefont {Ikeda},
  \citenamefont {Nomura}, \citenamefont {Matsuda}, \citenamefont {Tani},
  \citenamefont {Kobayashi}, \citenamefont {Watanabe},\ and\ \citenamefont
  {Sato}}]{IkedaRSI2017}%
  \BibitemOpen
  \bibfield  {author} {\bibinfo {author} {\bibfnamefont {A.}~\bibnamefont
  {Ikeda}}, \bibinfo {author} {\bibfnamefont {T.}~\bibnamefont {Nomura}},
  \bibinfo {author} {\bibfnamefont {Y.~H.}\ \bibnamefont {Matsuda}}, \bibinfo
  {author} {\bibfnamefont {S.}~\bibnamefont {Tani}}, \bibinfo {author}
  {\bibfnamefont {Y.}~\bibnamefont {Kobayashi}}, \bibinfo {author}
  {\bibfnamefont {H.}~\bibnamefont {Watanabe}}, \ and\ \bibinfo {author}
  {\bibfnamefont {K.}~\bibnamefont {Sato}},\ }\href {\doibase
  10.1063/1.4999452} {\bibfield  {journal} {\bibinfo  {journal} {Rev. Sci.
  Instrum.}\ }\textbf {\bibinfo {volume} {88}},\ \bibinfo {pages} {083906}
  (\bibinfo {year} {2017})}\BibitemShut {NoStop}%
\bibitem [{\citenamefont {Ikeda}\ \emph {et~al.}(2020)\citenamefont {Ikeda},
  \citenamefont {Matsuda},\ and\ \citenamefont {Sato}}]{IkedaPRL2020}%
  \BibitemOpen
  \bibfield  {author} {\bibinfo {author} {\bibfnamefont {A.}~\bibnamefont
  {Ikeda}}, \bibinfo {author} {\bibfnamefont {Y.~H.}\ \bibnamefont {Matsuda}},
  \ and\ \bibinfo {author} {\bibfnamefont {K.}~\bibnamefont {Sato}},\ }\href
  {\doibase 10.1103/PhysRevLett.125.177202} {\bibfield  {journal} {\bibinfo
  {journal} {Phys. Rev. Lett.}\ }\textbf {\bibinfo {volume} {125}},\ \bibinfo
  {pages} {177202} (\bibinfo {year} {2020})}\BibitemShut {NoStop}%
\bibitem [{\citenamefont {Noguchi}\ \emph {et~al.}(2002)\citenamefont
  {Noguchi}, \citenamefont {Kawamata}, \citenamefont {Okuda}, \citenamefont
  {Nojiri},\ and\ \citenamefont {Motokawa}}]{NoguchiPRB2002}%
  \BibitemOpen
  \bibfield  {author} {\bibinfo {author} {\bibfnamefont {S.}~\bibnamefont
  {Noguchi}}, \bibinfo {author} {\bibfnamefont {S.}~\bibnamefont {Kawamata}},
  \bibinfo {author} {\bibfnamefont {K.}~\bibnamefont {Okuda}}, \bibinfo
  {author} {\bibfnamefont {H.}~\bibnamefont {Nojiri}}, \ and\ \bibinfo {author}
  {\bibfnamefont {M.}~\bibnamefont {Motokawa}},\ }\href {\doibase
  10.1103/PhysRevB.66.094404} {\bibfield  {journal} {\bibinfo  {journal} {Phys.
  Rev. B}\ }\textbf {\bibinfo {volume} {66}},\ \bibinfo {pages} {094404}
  (\bibinfo {year} {2002})}\BibitemShut {NoStop}%
\bibitem [{\citenamefont {Sin~Naing}\ \emph {et~al.}(2006)\citenamefont
  {Sin~Naing}, \citenamefont {Kobayashi}, \citenamefont {Kobayashi},
  \citenamefont {Suzuki},\ and\ \citenamefont {Asai}}]{NaingJPSJ2006}%
  \BibitemOpen
  \bibfield  {author} {\bibinfo {author} {\bibfnamefont {T.}~\bibnamefont
  {Sin~Naing}}, \bibinfo {author} {\bibfnamefont {T.}~\bibnamefont
  {Kobayashi}}, \bibinfo {author} {\bibfnamefont {Y.}~\bibnamefont
  {Kobayashi}}, \bibinfo {author} {\bibfnamefont {M.}~\bibnamefont {Suzuki}}, \
  and\ \bibinfo {author} {\bibfnamefont {K.}~\bibnamefont {Asai}},\ }\href
  {\doibase 10.1143/JPSJ.75.084601} {\bibfield  {journal} {\bibinfo  {journal}
  {J. Phys. Soc. Jpn.}\ }\textbf {\bibinfo {volume} {75}},\ \bibinfo {pages}
  {084601} (\bibinfo {year} {2006})}\BibitemShut {NoStop}%
\bibitem [{\citenamefont {Hariki}\ \emph {et~al.}(2020)\citenamefont {Hariki},
  \citenamefont {Wang}, \citenamefont {Sotnikov}, \citenamefont {Tomiyasu},
  \citenamefont {Betto}, \citenamefont {Brookes}, \citenamefont {Uemura},
  \citenamefont {Ghiasi}, \citenamefont {Groot},\ and\ \citenamefont
  {Kun\v{e}s}}]{HarikiPRB2020}%
  \BibitemOpen
  \bibfield  {author} {\bibinfo {author} {\bibfnamefont {A.}~\bibnamefont
  {Hariki}}, \bibinfo {author} {\bibfnamefont {R.-P.}\ \bibnamefont {Wang}},
  \bibinfo {author} {\bibfnamefont {A.}~\bibnamefont {Sotnikov}}, \bibinfo
  {author} {\bibfnamefont {K.}~\bibnamefont {Tomiyasu}}, \bibinfo {author}
  {\bibfnamefont {D.}~\bibnamefont {Betto}}, \bibinfo {author} {\bibfnamefont
  {N.~B.}\ \bibnamefont {Brookes}}, \bibinfo {author} {\bibfnamefont
  {Y.}~\bibnamefont {Uemura}}, \bibinfo {author} {\bibfnamefont
  {M.}~\bibnamefont {Ghiasi}}, \bibinfo {author} {\bibfnamefont {F.~M. F.~d.}\
  \bibnamefont {Groot}}, \ and\ \bibinfo {author} {\bibfnamefont
  {J.}~\bibnamefont {Kun\v{e}s}},\ }\href {\doibase
  10.1103/PhysRevB.101.245162} {\bibfield  {journal} {\bibinfo  {journal}
  {Phys. Rev. B}\ }\textbf {\bibinfo {volume} {101}},\ \bibinfo {pages}
  {245162} (\bibinfo {year} {2020})}\BibitemShut {NoStop}%
\bibitem [{\citenamefont {Nasu}\ \emph {et~al.}(2016)\citenamefont {Nasu},
  \citenamefont {Watanabe}, \citenamefont {Naka},\ and\ \citenamefont
  {Ishihara}}]{NasuPRB2016}%
  \BibitemOpen
  \bibfield  {author} {\bibinfo {author} {\bibfnamefont {J.}~\bibnamefont
  {Nasu}}, \bibinfo {author} {\bibfnamefont {T.}~\bibnamefont {Watanabe}},
  \bibinfo {author} {\bibfnamefont {M.}~\bibnamefont {Naka}}, \ and\ \bibinfo
  {author} {\bibfnamefont {S.}~\bibnamefont {Ishihara}},\ }\href {\doibase
  10.1103/PhysRevB.93.205136} {\bibfield  {journal} {\bibinfo  {journal} {Phys.
  Rev. B}\ }\textbf {\bibinfo {volume} {93}},\ \bibinfo {pages} {205136}
  (\bibinfo {year} {2016})}\BibitemShut {NoStop}%
\bibitem [{\citenamefont {Altarawneh}\ \emph {et~al.}(2012)\citenamefont
  {Altarawneh}, \citenamefont {Chern}, \citenamefont {Harrison}, \citenamefont
  {Batista}, \citenamefont {Uchida}, \citenamefont {Jaime}, \citenamefont
  {Rickel}, \citenamefont {Crooker}, \citenamefont {Mielke}, \citenamefont
  {Betts}, \citenamefont {Mitchell},\ and\ \citenamefont {Hoch}}]{MoazPRL2012}%
  \BibitemOpen
  \bibfield  {author} {\bibinfo {author} {\bibfnamefont {M.~M.}\ \bibnamefont
  {Altarawneh}}, \bibinfo {author} {\bibfnamefont {G.~W.}\ \bibnamefont
  {Chern}}, \bibinfo {author} {\bibfnamefont {N.}~\bibnamefont {Harrison}},
  \bibinfo {author} {\bibfnamefont {C.~D.}\ \bibnamefont {Batista}}, \bibinfo
  {author} {\bibfnamefont {A.}~\bibnamefont {Uchida}}, \bibinfo {author}
  {\bibfnamefont {M.}~\bibnamefont {Jaime}}, \bibinfo {author} {\bibfnamefont
  {D.~G.}\ \bibnamefont {Rickel}}, \bibinfo {author} {\bibfnamefont {S.~A.}\
  \bibnamefont {Crooker}}, \bibinfo {author} {\bibfnamefont {C.~H.}\
  \bibnamefont {Mielke}}, \bibinfo {author} {\bibfnamefont {J.~B.}\
  \bibnamefont {Betts}}, \bibinfo {author} {\bibfnamefont {J.~F.}\ \bibnamefont
  {Mitchell}}, \ and\ \bibinfo {author} {\bibfnamefont {M.~J.~R.}\ \bibnamefont
  {Hoch}},\ }\href {\doibase 10.1103/PhysRevLett.109.037201} {\bibfield
  {journal} {\bibinfo  {journal} {Phys. Rev. Lett.}\ }\textbf {\bibinfo
  {volume} {109}},\ \bibinfo {pages} {037201} (\bibinfo {year}
  {2012})}\BibitemShut {NoStop}%
\bibitem [{\citenamefont {Ikeda}\ \emph {et~al.}(2016)\citenamefont {Ikeda},
  \citenamefont {Nomura}, \citenamefont {Matsuda}, \citenamefont {Matsuo},
  \citenamefont {Kindo},\ and\ \citenamefont {Sato}}]{IkedaLCO2016}%
  \BibitemOpen
  \bibfield  {author} {\bibinfo {author} {\bibfnamefont {A.}~\bibnamefont
  {Ikeda}}, \bibinfo {author} {\bibfnamefont {T.}~\bibnamefont {Nomura}},
  \bibinfo {author} {\bibfnamefont {Y.~H.}\ \bibnamefont {Matsuda}}, \bibinfo
  {author} {\bibfnamefont {A.}~\bibnamefont {Matsuo}}, \bibinfo {author}
  {\bibfnamefont {K.}~\bibnamefont {Kindo}}, \ and\ \bibinfo {author}
  {\bibfnamefont {K.}~\bibnamefont {Sato}},\ }\href {\doibase
  10.1103/PhysRevB.93.220401} {\bibfield  {journal} {\bibinfo  {journal} {Phys.
  Rev. B}\ }\textbf {\bibinfo {volume} {93}},\ \bibinfo {pages} {220401(R)}
  (\bibinfo {year} {2016})}\BibitemShut {NoStop}%
\bibitem [{\citenamefont {Haverkort}\ \emph {et~al.}(2006)\citenamefont
  {Haverkort}, \citenamefont {Hu}, \citenamefont {Cezar}, \citenamefont
  {Burnus}, \citenamefont {Hartmann}, \citenamefont {Reuther}, \citenamefont
  {Zobel}, \citenamefont {Lorenz}, \citenamefont {Tanaka}, \citenamefont
  {Brookes}, \citenamefont {Hsieh}, \citenamefont {Lin}, \citenamefont {Chen},\
  and\ \citenamefont {Tjeng}}]{Haverkort}%
  \BibitemOpen
  \bibfield  {author} {\bibinfo {author} {\bibfnamefont {M.~W.}\ \bibnamefont
  {Haverkort}}, \bibinfo {author} {\bibfnamefont {Z.}~\bibnamefont {Hu}},
  \bibinfo {author} {\bibfnamefont {J.~C.}\ \bibnamefont {Cezar}}, \bibinfo
  {author} {\bibfnamefont {T.}~\bibnamefont {Burnus}}, \bibinfo {author}
  {\bibfnamefont {H.}~\bibnamefont {Hartmann}}, \bibinfo {author}
  {\bibfnamefont {M.}~\bibnamefont {Reuther}}, \bibinfo {author} {\bibfnamefont
  {C.}~\bibnamefont {Zobel}}, \bibinfo {author} {\bibfnamefont
  {T.}~\bibnamefont {Lorenz}}, \bibinfo {author} {\bibfnamefont
  {A.}~\bibnamefont {Tanaka}}, \bibinfo {author} {\bibfnamefont {N.~B.}\
  \bibnamefont {Brookes}}, \bibinfo {author} {\bibfnamefont {H.~H.}\
  \bibnamefont {Hsieh}}, \bibinfo {author} {\bibfnamefont {H.~J.}\ \bibnamefont
  {Lin}}, \bibinfo {author} {\bibfnamefont {C.~T.}\ \bibnamefont {Chen}}, \
  and\ \bibinfo {author} {\bibfnamefont {L.~H.}\ \bibnamefont {Tjeng}},\ }\href
  {\doibase 10.1103/PhysRevLett.97.176405} {\bibfield  {journal} {\bibinfo
  {journal} {Phys. Rev. Lett.}\ }\textbf {\bibinfo {volume} {97}},\ \bibinfo
  {pages} {176405} (\bibinfo {year} {2006})}\BibitemShut {NoStop}%
\bibitem [{\citenamefont {Radaelli}\ and\ \citenamefont
  {Cheong}(2002)}]{Radaelli}%
  \BibitemOpen
  \bibfield  {author} {\bibinfo {author} {\bibfnamefont {P.~G.}\ \bibnamefont
  {Radaelli}}\ and\ \bibinfo {author} {\bibfnamefont {S.~W.}\ \bibnamefont
  {Cheong}},\ }\href {\doibase 10.1103/PhysRevB.66.094408} {\bibfield
  {journal} {\bibinfo  {journal} {Phys. Rev. B}\ }\textbf {\bibinfo {volume}
  {66}},\ \bibinfo {pages} {094408} (\bibinfo {year} {2002})}\BibitemShut
  {NoStop}%
\bibitem [{\citenamefont {Ikeda}\ \emph {et~al.}(2022)\citenamefont {Ikeda},
  \citenamefont {Matsuda}, \citenamefont {Zhou}, \citenamefont {Peng},
  \citenamefont {Ishii}, \citenamefont {Yajima}, \citenamefont {Kubota},
  \citenamefont {Inoue}, \citenamefont {Inubushi}, \citenamefont {Tono},\ and\
  \citenamefont {Yabashi}}]{IkedaAPL2022}%
  \BibitemOpen
  \bibfield  {author} {\bibinfo {author} {\bibfnamefont {A.}~\bibnamefont
  {Ikeda}}, \bibinfo {author} {\bibfnamefont {Y.~H.}\ \bibnamefont {Matsuda}},
  \bibinfo {author} {\bibfnamefont {X.}~\bibnamefont {Zhou}}, \bibinfo {author}
  {\bibfnamefont {S.}~\bibnamefont {Peng}}, \bibinfo {author} {\bibfnamefont
  {Y.}~\bibnamefont {Ishii}}, \bibinfo {author} {\bibfnamefont
  {T.}~\bibnamefont {Yajima}}, \bibinfo {author} {\bibfnamefont
  {Y.}~\bibnamefont {Kubota}}, \bibinfo {author} {\bibfnamefont
  {I.}~\bibnamefont {Inoue}}, \bibinfo {author} {\bibfnamefont
  {Y.}~\bibnamefont {Inubushi}}, \bibinfo {author} {\bibfnamefont
  {K.}~\bibnamefont {Tono}}, \ and\ \bibinfo {author} {\bibfnamefont
  {M.}~\bibnamefont {Yabashi}},\ }\href {\doibase 10.1063/5.0088134} {\bibfield
   {journal} {\bibinfo  {journal} {Appl. Phys. Lett.}\ }\textbf {\bibinfo
  {volume} {120}},\ \bibinfo {pages} {142403} (\bibinfo {year}
  {2022})}\BibitemShut {NoStop}%
\end{thebibliography}%

\paragraph*{\rm{\textbf{Acknowledgements}}}
AI is financially supported by MEXT Leading Initiative for Excellent Young Researchers Grant No.~JPMXS0320210021, JSPS KAKENHI Grant-in-Aid for Early-Career Scientists Grant No.~18K13493, and Basic Science Program No.~18-001 of Tokyo Electric Power Company (TEPCO) memorial foundation.

\paragraph*{\rm{\textbf{Author contributions}}}
A.I. designed the research.
K.S prepared the sample.
A.I., Y.H.M., H.S., Y. I., D.N., S.T. performed the experiments.
A.I. analyzed the data.
J.N. and A.I. developed the model.
J.N. performed the calculations.
A.I, Y.H.M., J.N. discussed the results.
A.I. wrote the manuscript with input from all co-authors.

\paragraph*{\rm{\textbf{Competing interests}}}
The authors declare no competing interests.

\end{document}

\title{Supplementary Information}%LaCoO$_{3}$ 
\author{Akihiko~Ikeda}
\affiliation{Institute for Solid State Physics, University of Tokyo, Kashiwa, Chiba 277-8581, Japan}
\affiliation{Department of Engineering Science, University of Electro-Communications, Chofu, Tokyo 182-8585, Japan}
\author{Yasuhiro~H.~Matsuda}
\affiliation{Institute for Solid State Physics, University of Tokyo, Kashiwa, Chiba 277-8581, Japan}
\author{Keisuke~Sato}
\affiliation{National Institute of Technology, Ibaraki College, Hitachinaka, Ibaraki 312-0011, Japan}
\author{Yuto~Ishii}
\author{Hironobu~Sawabe}
\author{Daisuke~Nakamura}
\author{Shojiro~Takeyama}
\affiliation{Institute for Solid State Physics, University of Tokyo, Kashiwa, Chiba 277-8581, Japan}
\author{Joji~Nasu}
\affiliation{Department of Physics, Tohoku University, Sendai, Miyagi 980-8578, Japan}
\affiliation{PRESTO, Japan Science and Technology Agency, Honcho Kawaguchi, Saitama 332-0012, Japan}

\date{\today}

\maketitle

\section{Supplementary Note 1: Experimental details}
\subsection{Generation of 600 T using electromagnetic flux compression method in ISSP}

 \begin{figure}[h]
 \includegraphics[width = 1.0\linewidth]{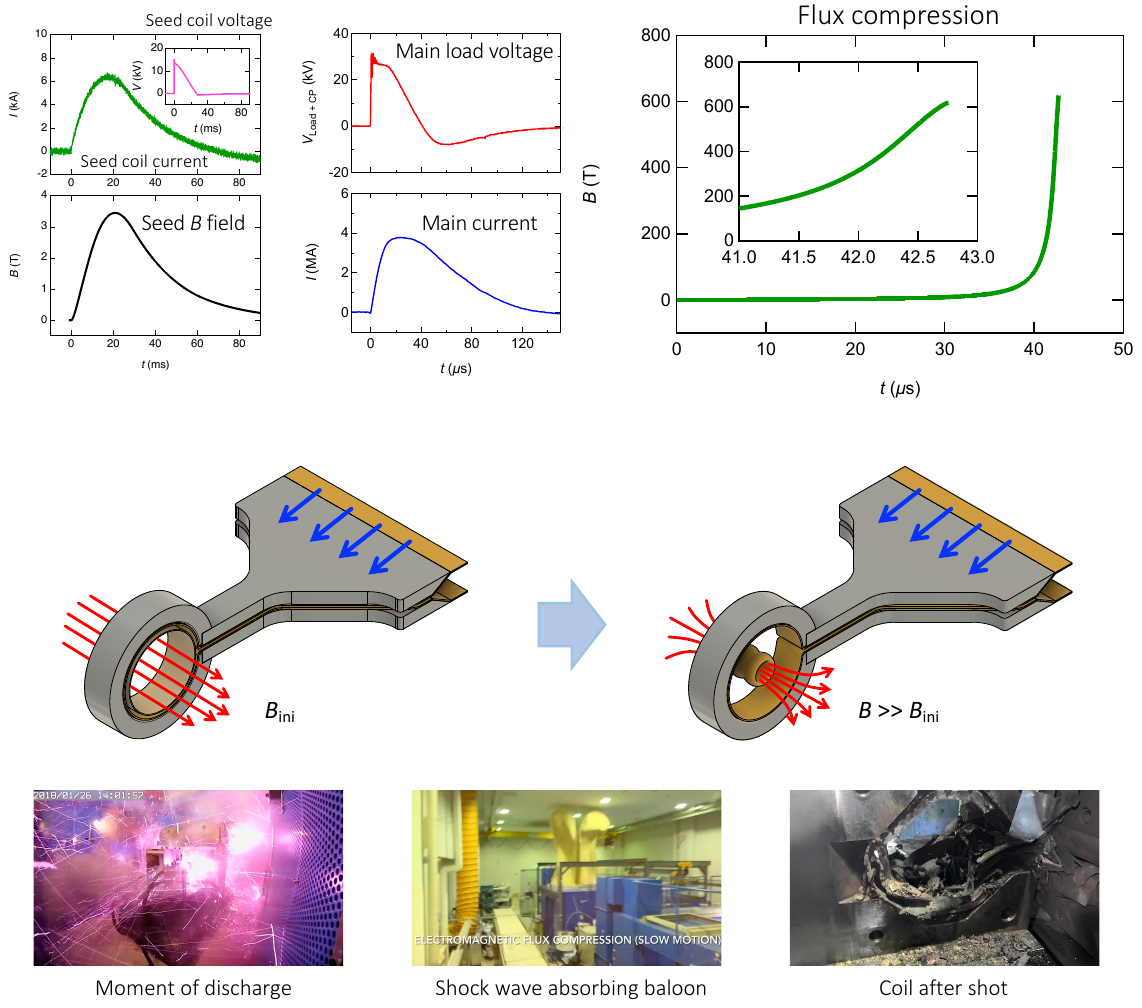}%
 \caption{The seed field generation, the main current used for the flux compression, and the typical magnetic field waveform in the present EMFC experiments.
 \label{fig01s}}
 \end{figure}

The electromagnetic flux compression (EMFC) method is used to generate an ultrahigh magnetic field of 600 T in ISSP, University of Tokyo, that we implemented recently \cite{NakamuraRSI2018}. 
In the EMFC method, the ultrahigh magnetic field is generated by concentrating the seed magnetic flux into a small cross-section and volume.
A cylindrical metallic liner is rapidly compressed using electromagnetic force for the flux concentration.
The electromagnetic force originates in the current repulsion between the main coil and the induced current inside the metallic liner.
A Helmholtz coil is used to generate a seed magnetic field of 3.4 T, where the energy stored in the sub capacitor bank is 0.95 MJ at a charging voltage of 13.8 kV.
The Peak current of the main coil is 3.8 MA discharged with a charging voltage of 40 kV, where the energy stored in the main capacitor bank is 2.56 MJ.
In Supplementary Figs. \ref{fig01s}, the waveform of the seed magnetic field, main current and magnetic field are shown along with the schematic drawing of the flux compression and the pictures of the EMFC experiment in ISSP.

  \begin{figure}[t]
 \includegraphics[width = 1.0\linewidth]{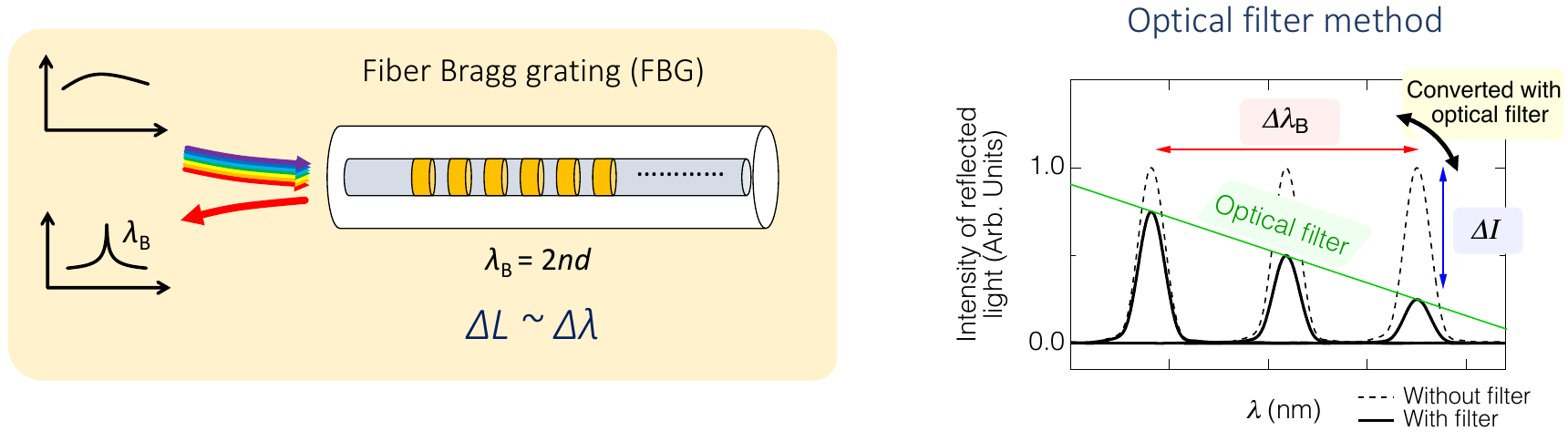}
 \caption{Schematic drawing of the principle of FBG strain gauge and optical filter method for 100 MHz high-speed detection of strain.
 \label{fig02s}}
 \end{figure}

  \begin{figure}[b]
 \includegraphics[width = 1.0\linewidth]{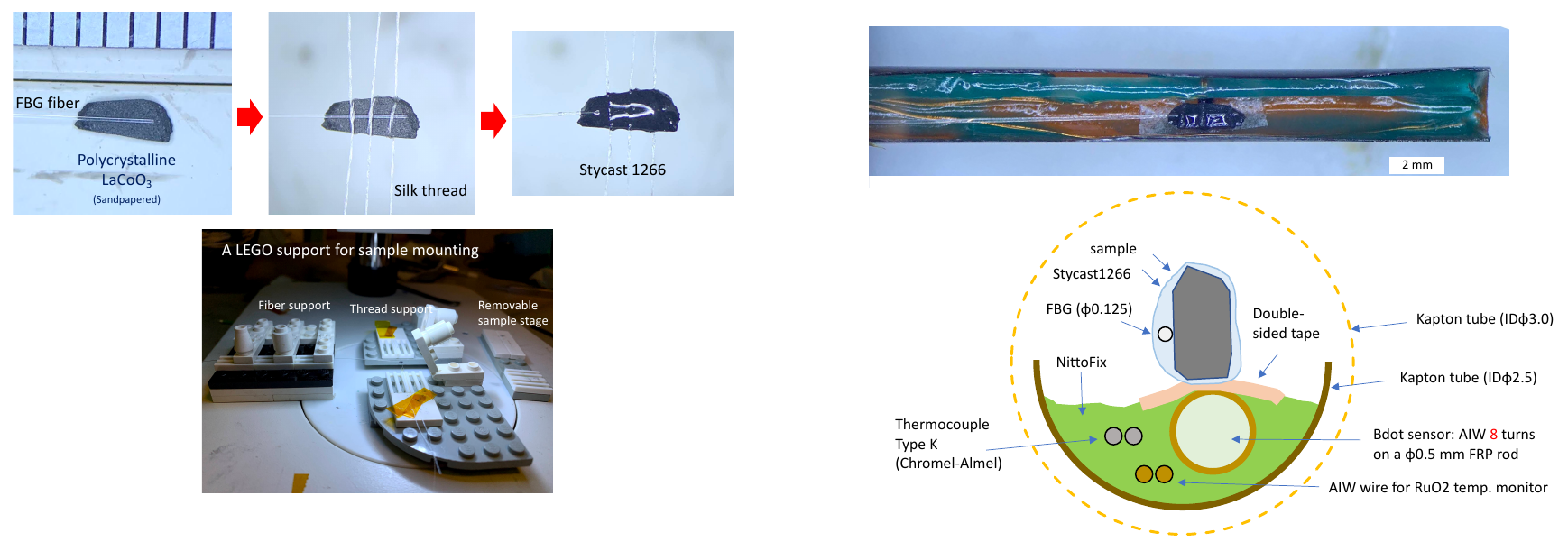}
 \caption{Technique for gluing the sample and FBG fiber together using silk thread. A LEGO(R) support was used. The picture and the schematic cross-section of the sample probe are shown.
 \label{fig03s}}
 \end{figure}

\subsection{High-speed magnetostriction measurement using fiber Bragg grating and optical filter method for the use up to 600 T }

   \begin{figure}
 \includegraphics[width = 1\linewidth]{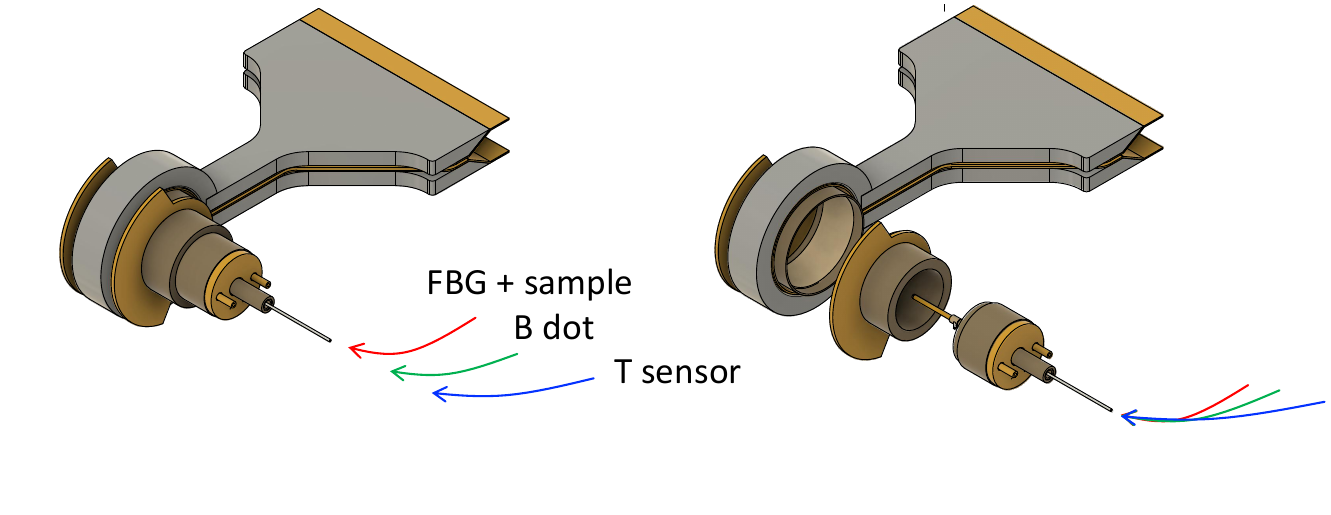}
 \caption{The image of the vacuum chamber and the cryostat for the EMFC experiments. The sample and all the sensors are inserted from the right end of the image, being removable from the system. 
 \label{fig04s}}
 \end{figure}

We have employed the original high-speed strain gauge utilizing the fiber Bragg grating (FBG) technique and optical filter method  \cite{IkedaRSI2017, IkedaPRL2020} to measure the longitudinal magnetostriction of LaCoO$_{3}$ under ultrahigh magnetic fields generated using EMFC.
Requirements for the magnetostriction measurement with EMFC are 100 MHz high speed and noiselessness.
FBG is an optical fiber equipped with optical Bragg grating in the core, that we can use as an optical strain gauge as schematically shown in Supplementary Fig. \ref{fig02s}.
The strain of the optical fiber sensitively appears as the shift of the Bragg wavelength.
Thus, the shift of the Bragg wavelength can be monitored optically from a distance, which is of benefit considering the large noise from the explosion and electric discharge involved in the EMFC technique.
We used the optical filter method as the detection scheme \cite{IkedaRSI2017}.
This is because the required measurement speed is as high as 100 MHz in the EMFC experiment.
In the optical filter method, the shift of the Bragg wavelength is converted to the change of the intensity of the optical signal as schematically shown in Supplementary Fig. \ref{fig02s}.
The change in the intensity of the optical signal is monitored by the InGaAs avalanche photodiode.

The strain of the sample is transmitted to the FBG fiber that is glued together [See Supplementary Fig. \ref{fig03s}].
The gluing has been carried out in the following manner.
 First, two or three silk threads are used to tie the sample to the FBG fiber.
 This is performed on a base made from LEGO(R) bricks.
 The bottom base is removed and STYCAST 1266 is used to glue the sample and the FBG fiber with the silk thread.
 The silk thread is used to make sure the FBG fiber is intact with the sample during the drying proccesse.
The gluing method is the heart of the technique. 
 The glued sample and FBG are placed on a Kapton sample holder tube.
 The induction type magnetic field sensor and temperature monitors of thermocouple or resistive RuO$_{2}$ tip are placed nearby in diameter of $\phi$ 2.5 mm [See Supplementary Fig. \ref{fig03s}].
 The probe equipped with the sample, FBG strain gauge, magnetic field sensor, and temperature sensors is inserted into a He flow type cryostat.
 The cryostat is placed at the field center inside a vacuum chamber made of bakelite parts, which serve as a thermal insulator and are also advantageous for the Cu liner implosion in the EMFC technique [See Supplementary Fig. \ref{fig04s}].
 
   \begin{figure}
 \includegraphics[width = 0.9\linewidth]{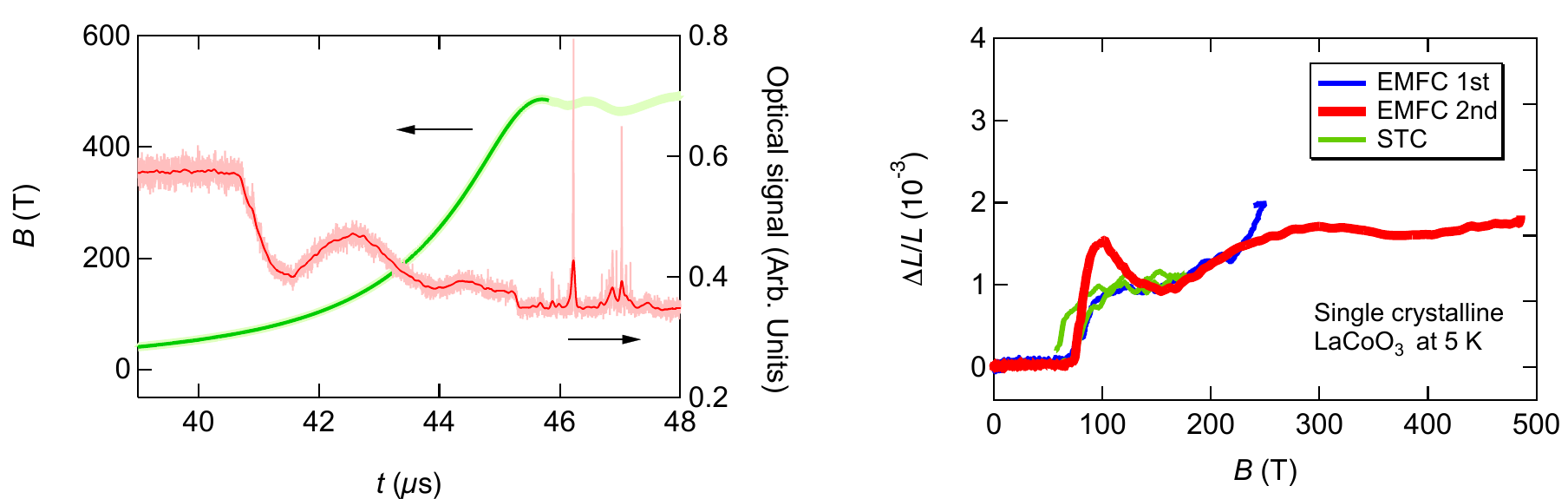}
 \caption{Results of magnetostriction measurement using single crystalline LaCoO$_{3}$ up to 600 T at 5 K.
 \label{fig05s}}
 \end{figure}

  \subsection{Magnetostriction measurements with single crystalline samples}

Single crystalline samples are measured up to 200 T as shown in Supplementary Fig. \ref{fig05s}.
It is difficult to observe the temperature dependence coherently with many pulses because single crystalline samples break themselves due to the field-induced phase transition taking place too rapidly.
The strong vibrations of the data are present  [See Supplementary Fig. \ref{fig05s}].
The temperature-dependent data is disturbed and incomparable due to the re-gluing fresh samples to new optical fibers.
On the other hand, polycrystalline samples are robust against the spin-state transitions triggered by the rapid magnetic field pulse.
Poly-and single-crystalline samples show basically identical results with minor differences in the sharpness of the spin-state transitions.

 \subsection{Consideration on spin-orbit interactions}

We comment on the difference between $d V/V$ and $d L/L$.
Spin-state originates purely in the electron spin ($\bm{S}$) when no spin-orbit interaction ($\bm{S}\cdot \bm{L}$) is in play.
In that case, $dL/L$ can represent $dV/V$.
Actually, such a situation is common in transition metal oxides, where the orbital angular momentum ($\bm{L}$) is usually quenched.
In the case of LaCoO$_{3}$, the orbital angular momentum is not completely quenched.
Thus, an influence of spin-orbit interaction may result in $d V/V\neq d L/L$ in magnetostriction even for the polycrystalline sample used in the present study.
In reality we may ignore this because it is reported small enough to be neglected in the present experiment by an electron resonance study showing the almost isotropic g-factors ($g_{\parallel}$ = 3.35, $g_{\perp}$ = 3.55) in LaCoO$_{3}$ \cite{NoguchiPRB2002}.

We comment on another experimental fact indicating that we can ignore the spin-orbit couplings.
Previously, we argued that the magnetization increases with increasing magnetic fields in $\beta$ phase although magnetostriction is a constant, which may induce the spin-driven lattice change through spin-orbit coupling.
We confirmed that this is not the case with measurements of magnetostriction and magnetization at room temperature, where only magnetization increases up to 50 T without any increase in magnetostriction  \cite{IkedaPRL2020}.
It is reasonable to assume that the spin-orbit coupling, a local interaction, is temperature independent and is also absent at 78 and 108 K.
This justifies our assumption that the longitudinal magnetostriction is proportional to lattice volume in the present study.
The influence of orbital order is also unlikely either because we presently used a polycrystalline sample.
 
\subsection{Effect of mechanical vibrations}

We exclude the possibility of mechanical vibration as a source of the oscillating feature in 108 K data [see Fig. 3c in the main text] on the following basis.
The key idea is temperature dependence. If the oscillating feature in 108 K data is actually a vibration, it should have appeared in the 78 K data with similar amplitude and frequency. We see only a small oscillating feature in the 78 K data above $B_{\rm c2}$. The sharp slope features in 78 K are more apparent. 108 K is only 30 K above 78 K. The mechanical property is similar in this temperature change as reported in an ultrasound study \cite{NaingJPSJ2006}. Similarly, at the data 185 K, no negative slope is observed. This also indicates that the negative slopes of the data at 78 K and 108 K are intrinsic features.

Quantitatively, If the feature in 108 K is vibration, the shock propagation speed corresponds to 1.5 km/s with 250 kHz vibration and a sample size of 3 mm. In the same way, we obtain 3 km/s with 600 kHz vibration and a sample size of 3 mm. According to Ref. \cite{NaingJPSJ2006}, the sound velocity change from 5 K to 108 K is 15 \%. Thus the oscillating feature in 108 K data is too slow for a shock propagation vibration.

\section{Supplementary Note 2: Mean field calculation}

We present the details of the theoretical model for the magnetic properties of trivalent cobalt ions ($3d^6$) in LaCoO$_3$ in an applied magnetic field and introduce the calculation method to address the model.
We also show the results and discussion.

\subsection{Model}

\begin{figure}[t]
\begin{center}
\includegraphics[width=0.6\columnwidth,clip]{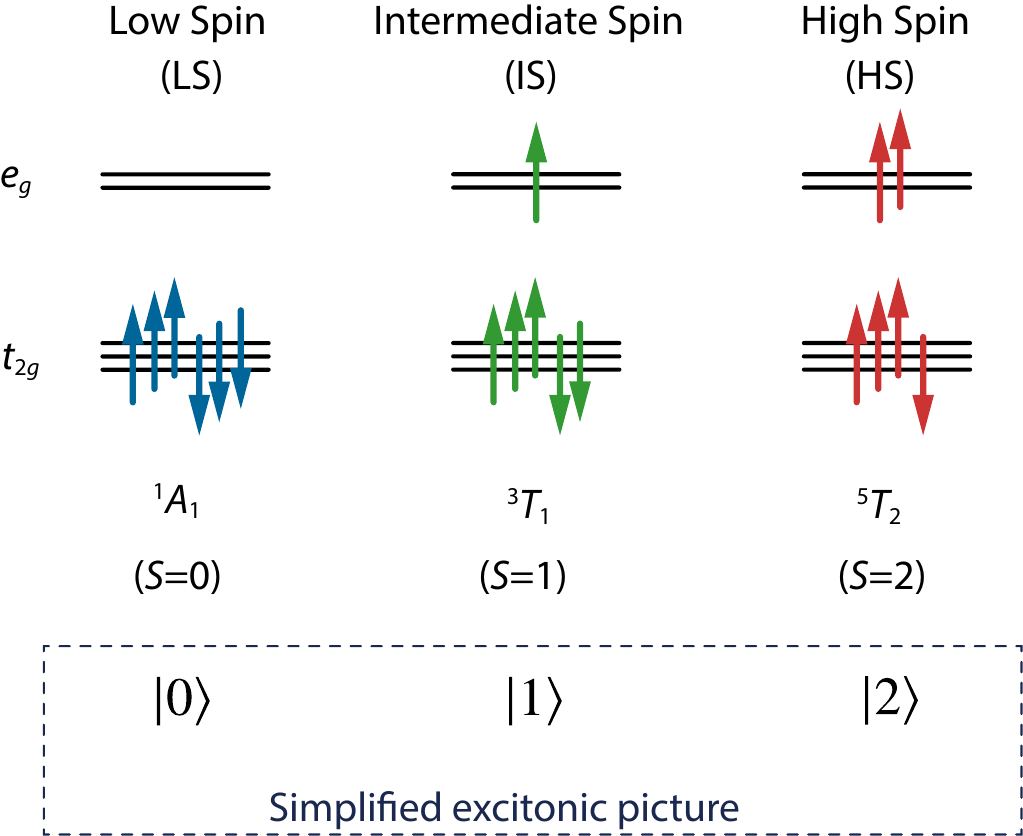}
\caption{
Schematic picture of the three spin states and their excitonic representations simplified in the magnetic field.
}
\label{fig:spinstates}
\end{center}
\end{figure}

We start from the five-orbital Hubbard model under the magnetic field to understand the properties of $3d$ electrons in cobalt ions.
This model is composed of the following four terms: the transfer integrals between $3d$ orbitals in neighboring ions, $\mathcal{H}_t$, on-site Coulomb interactions $\mathcal{H}_U$, crystalline electric field $\mathcal{H}_{\rm crys}$ with the octahedral symmetry, and Zeeman term $\mathcal{H}_{Z}$.
Since there are five orbitals, which are occupied by six electrons per site, the Hamiltonian is extremely complicated.
The competition between the local Hamiltonians, $\mathcal{H}_U$ and $\mathcal{H}_{\rm crys}$, is understood from the well-known Tanabe-Sugano diagram.
In the $d^6$ configurations, three types of spin states, low-spin (LS), intermediate spin (IS), and high-spin states (HS) shown in Supplementary Fig.~\ref{fig:spinstates} play a crucial role in the magnetic properties of a local cobalt site. 
When the on-site Coulomb interactions are only taken into account, the ground state of the local $d^6$ configurations is the $LS$ multiplet ${}^5 D$ ($S=2, L=2$) corresponding to the HS state while the IS and LS are excited states.
By introducing the crystalline electric field of the octahedral ligand arrangement, the ground state changes to the LS state where the $t_{2g}$ orbitals are fully occupied.
The IS can be a first excited state but does not become a ground state within a single ion system.
Nevertheless, intensive experimental and theoretical studies on cobalt oxides suggested the possibility of the emergence of the IS state due to the competition between itinerant and localized natures of electrons.

Next, we discuss the effect of the Zeeman term $\mathcal{H}_{Z}$.
The strong magnetic field lifts spin and orbital degeneracies, which allows us to extract a state with the Zeeman energy minimized in each spin-state manifold.
Such states in LS, IS, and HS manifold are referred to as $\ket{0}$, $\ket{1}$, and $\ket{2}$, respectively.
Here, we neglect the degeneracy surviving under the magnetic field in each spin state for simplicity.
We construct the low-energy effective Hamiltonian in the subspace composed of the direct products of the three local states $\ket{0}$, $\ket{1}$, and $\ket{2}$ in the strong correlation limit.
The effective Hamiltonian is written as
\begin{align}
    \mathcal{H} = \mathcal{H}_{\rm{loc}} 
    + \mathcal{H}_{\rm{int}} + \mathcal{H}_{\rm{trans}} + \mathcal{H}_{\rm{dual}}.\label{ham}
\end{align}
The first term $\mathcal{H}_{\rm{loc}}$ represents the local energy levels of the IS and HS states, which is given by
\begin{align}
    \mathcal{H}_{\rm{loc}} = \left(E_{\rm{1}} - g_{\rm{1}}h\right)\sum_{i} n_{i}^{\rm{(1)}} + \left(E_{\rm{2}} - g_{2}h\right)\sum_{i} n_{i}^{\rm{(2)}},
\end{align}
where $E_{\rm{1}}$ and $E_{\rm{2}}$ are the on-site energies of $\ket{1}$ and $\ket{2}$ measured from the LS one, respectively, and $n_{i}^{(1)}=\ket{1}_i\bra{1}_i$ and $n_{i}^{(2)}=\ket{2}_i\bra{2}_i$ are the number operators at site $i$.
The $g$ factors for the IS and HS states are written as $g_{1}=2$ and $g_{2}=4$ under the magnetic field $h$.
$E_1$ and $E_2$ should be positive because the ground state without magnetic fields is LS state.

The other terms in Eq.~\eqref{ham} are deduced from the second-order perturbation process with respect to $\mathcal{H}_t$.
For simplicity, we neglect the orbital anisotropy related to bond-dependent transfer integrals in $\mathcal{H}_t$.
First, we consider the diagonal part, where a local spin state is not changed by the perturbation process.
The transfer integrals between the $e_g$ orbitals at neighboring sites is much larger than those involving the $t_{2g}$ orbitals when the neighboring octahedra share their corners.
In this case, the energy gain is yielded by the perturbation process for the IS-IS, LS-HS, LS-IS, and IS-HS configurations at neighboring sites.
While the IS state does not become the ground state in a single ion system, one expects more perturbation processes involving the IS state than the HS because of the orbital degeneracy of the $e_g$ orbitals in the IS state.
Thus, we extract the IS-IS and LS-HS configurations as dominant contributions, and the corresponding Hamiltonian is given by 
\begin{align}
    \mathcal{H}_{\rm{int}} =&  - J_{\rm{11}}\sum_{\braket{i j}} n_{i}^{\rm{(1)}}n_{j}^{\rm{(1)}} - J_{\rm{02}}\sum_{\braket{i j}} \big(n_{i}^{\rm{(2)}}n_{j}^{\rm{(0)}} + n_{i}^{\rm{(0)}}n_{j}^{\rm{(2)}}\big),
\end{align}
where $n_i^{(0)}$ is the number operator for the LS state.
Note that $J_{11}$ and $J_{02}$ should be positive because of the second-order perturbation.

\begin{figure}[t]
\begin{center}
\includegraphics[width=0.8\columnwidth,clip]{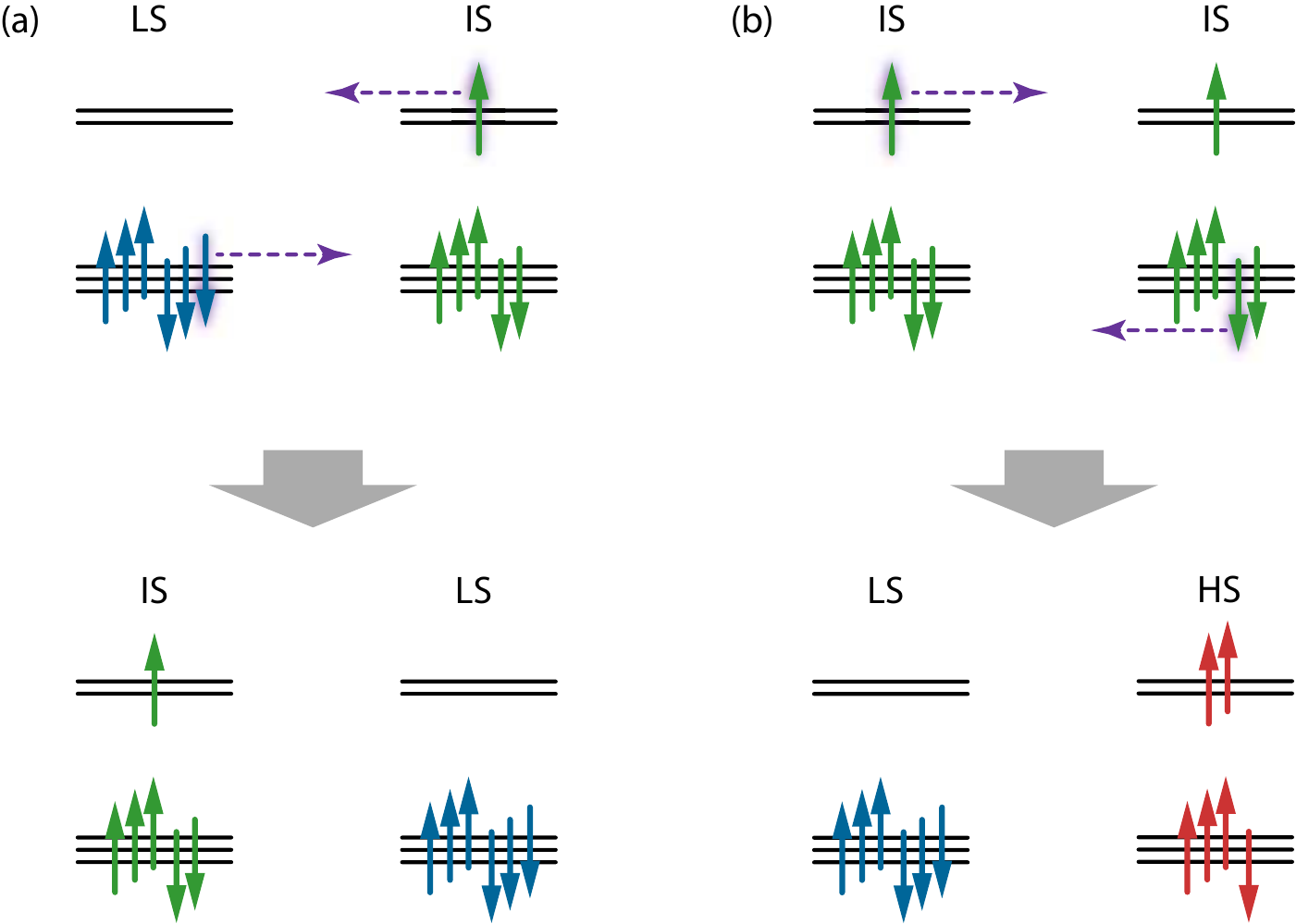}
\caption{
Two types of second-order perturbation processes changing neighboring local spin states:
(a) the exchange process between LS and IS states and (b) the process where two IS states are changed to LS and HS states, which is regarded as a fusion process of two single-excitons to a bi-exciton. 
}
\label{fig:exchange}
\end{center}
\end{figure}

Next, we consider the exchange terms in the perturbation expansion.
As suggested in Ref.~\onlinecite{HarikiPRB2020}, there are two dominant processes: the exchange process between neighboring LS and IS states and the change from the neighboring two IS states into LS and HS states. 
The former is interpreted as the hopping of a single exciton to a neighboring site, $\ket{0, 1}_{ij}\rightleftharpoons\ket{1, 0}_{ij}$, and the latter as the fusion of two single excitons into a bi-exciton (and the fission of it), $\ket{1, 1}_{ij}\rightleftharpoons\ket{0, 2}_{ij}$.
These contributions are given by
\begin{align}
\mathcal{H}_{\rm{trans}} =& -t \sum_{\braket{i j}} \left(\ket{0, 1}_{ij}\bra{1, 0}_{i j} + \ket{1, 0}_{ij}\bra{0, 1}_{ij}\right),\\
\mathcal{H}_{\rm{dual}} =& -V\sum_{\braket{i j}}\bigg[\ket{1, 1}_{ij} \left(\bra{2, 0}_{ij} + \bra{0, 2}_{ij} \right) +\left(\ket{2, 0}_{ij} + \ket{0, 2}_{ij} \right) \bra{1, 1}_{ij} \bigg],
\end{align}
where $\ket{k, k'}_{ij}=\ket{k}_i\otimes \ket{k'}_j$.

The exchange constants, $E_1$, $E_2$, $J_{11}$, $J_{02}$, $t$, and $V$ should be obtained from the perturbation procedure, but determining their values is a significantly tough problem.
To avoid this, we regard these constants as external parameters.
The signs of $t$ and $V$ cannot be determined directly.
Fortunately, these can be inverted by an appropriate gauge transformation in the case of the bipartite lattice.
Here, we consider the three-dimensional cubic lattice with the coordination number $z=6$.
In this case, the obtained phase diagram does not depend on the signs.
Thus, we assume that $t$ and $V$ are positive hereafter.

\subsection{Mean-field theory}

Here, we show the details of the mean-field theory applied to the model Hamiltonian given in Eq.~\eqref{ham}.
We introduce two types of pseudospins, $\bm{\tau}_i$ and $\bm{\rho}_i$ at site $i$, which are defined by
\begin{align}
    \tau_i^x = \ket{0}_i\bra{1}_i+ {\rm H.c.},\quad
    \tau_i^y = i\ket{0}_i\bra{1}_i+ {\rm H.c.},\\
    \rho_i^x = \ket{1}_i\bra{2}_i+ {\rm H.c.},\quad
    \rho_i^y = i\ket{1}_i\bra{2}_i+ {\rm H.c.},
\end{align}
respectively.
Using these pseudospins, $\mathcal{H}_{\rm{trans}}$ and $\mathcal{H}_{\rm{dual}}$ are rewritten as
\begin{align}
    \mathcal{H}_{\rm{trans}} =& -\frac{t}{2}\sum_{\braket{ij}}\left(\tau_i^x \tau_j^x+\tau_i^y \tau_j^y\right),\\
    \mathcal{H}_{\rm{dual}} =& -\frac{V}{2}\sum_{\braket{ij}}\left(\tau_i^x \rho_j^x+\rho_i^x \tau_j^x+\tau_i^y \rho_j^y+\rho_i^y \tau_j^y\right).
\end{align}
We apply the mean-field decoupling to $\mathcal{H}_{\rm{int}}$, $\mathcal{H}_{\rm{trans}}$, and $\mathcal{H}_{\rm{dual}}$, where two-sublattice orders are assumed.
The order parameters are given by $\braket{\tau^x}_X$, $\braket{\tau^y}_X$, $\braket{\rho^x}_X$, $\braket{\rho^y}_X$, and $\braket{n^k}_X$, where $k=0,1,2$ and $X(=A,B)$ is the sublattice index.
The operator $\bm{\tau}$ corresponds to a mixing between the LS and IS states, and a nonzero $\braket{\bm{\tau}}_X$ suggests excitonic condensation, while a nonzero expectation value of the operator $\bm{\rho}$ indicates a spontaneous hybridization between IS and HS states.
Note that the U(1) symmetry appears to be present in $\mathcal{H}_{\rm{trans}}$ in the pseudospin space but it is due to the simplification of the effective model.
If we carry out a more rigorous derivation of the effective model from the multi-orbital Hubbard model, this symmetry should be absent in the obtained Hamiltonian~\cite{NasuPRB2016}.

\subsection{Mean-field result}

\begin{figure}[t]
\begin{center}
\includegraphics[width=\columnwidth,clip]{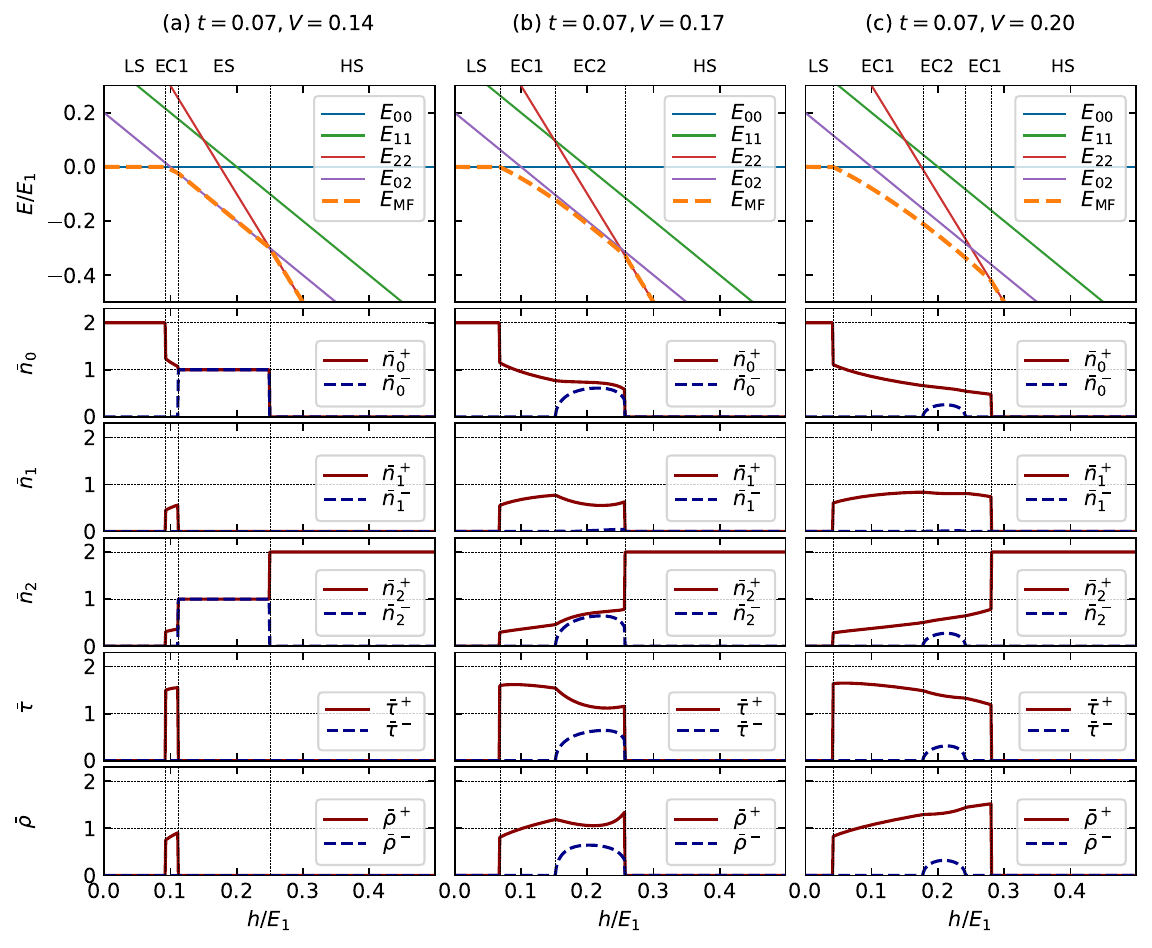}
\caption{
Mean-field results for the effective model given in Eq.~\eqref{ham} at (a) $V=0.14$, (b) $V=0.17$, and (c) $V=0.20$.
The other parameters are chosen to be ($E_1$, $E_2$, $J_{11}$, $J_{02}$, $t$) = (1.0, 0.7, 0.2, 0.05, 0.07).
The top panels represent the mean-field energy and classical energies given in Eq.~\eqref{eq:classical_energy} as functions of $h$.
The others show the field dependence of the mean-fields given in Eq.~\eqref{eq:mfs}.
The energy unit in the present calculations is $E_1$, which is estimated as $\sim 10$~meV correspinding to $\sim 100$~T.
}
\label{fig:mf}
\end{center}
\end{figure}

\begin{figure}[t]
\begin{center}
\includegraphics[width=0.8\columnwidth,clip]{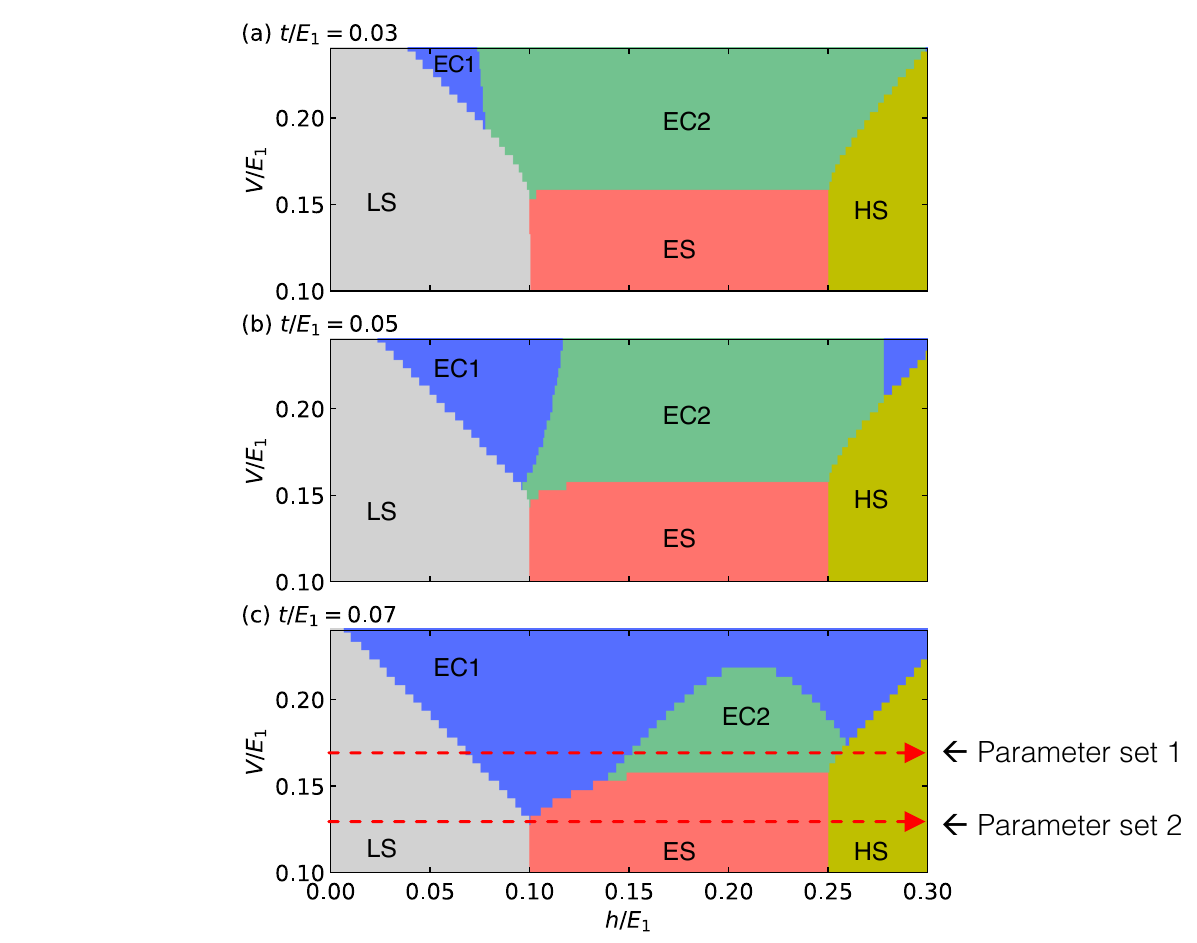}
\caption{\bl{
Phase diagram on the parameter space of $V/E_{1}$ and $h/E_{1}$ with the parameter variations of $t$ = (a) 0.03, (b) 0.05, and (c) 0.07.
The other parameters are chosen to be ($E_{1}$, $E_{2}$, $J_{11}$, $J_{02}$) =  (1.0, 0.7, 0.2, 0.05).
The horizontal dashed arrows indicate the parameter sets introduced in the main text.
}}
\label{fig:phase}
\end{center}
\end{figure}

We show the results of mean-field calculations for the effective model given in Eq.~\eqref{ham}.
Figure~\ref{fig:mf} shows the magnetic-field dependence of the mean-field energy, classical energies given by
\begin{align}
    \begin{cases}
        E_0=0,\\
        E_{11}=E_1-g_1 h -zJ_{11}/2,\\
        E_{22}=E_2-g_2 h,\\
        E_{02} = (E_2-g_2 h - z J_{02})/2,
    \end{cases}
    \label{eq:classical_energy}
\end{align}
and the order parameters given by
\begin{align}
    \begin{cases}
        \bar{n}_k^{\pm} = \left|\braket{n^{(k)}}_A\pm \braket{n^{(k)}}_B\right|, \qquad (k=0,1,2)\\
    \bar{\tau}^{\pm} = \sqrt{\left(\braket{\tau^x}_A\pm \braket{\tau^x}_B\right)^2 + \left(\braket{\tau^y}_A\pm \braket{\tau^y}_B\right)^2},\\
    \bar{\rho}^{\pm} = \sqrt{\left(\braket{\rho^x}_A\pm \braket{\rho^x}_B\right)^2 + \left(\braket{\rho^y}_A\pm \braket{\rho^y}_B\right)^2}.
\end{cases}
\label{eq:mfs}
\end{align}
We choose the parameters in Eq.~\eqref{ham} such that the LS is realized in the ground state in the absence of the magnetic field and this state changes to the LS-HS order and uniform HS state successively by applying the magnetic field in the classical case where $\mathcal{H}_{\rm{trans}}$ and $\mathcal{H}_{\rm{dual}}$ are dropped.
The LS-HS order is regarded as the exciton solid (ES).
The previous experimental studies suggest the situation~\cite{MoazPRL2012, IkedaLCO2016, IkedaPRL2020}, and thereby, we focus on the parameters shown in Supplementary Fig.~\ref{fig:mf}, which satisfy the above conditions.

The parameters in Supplementary Fig.~\ref{fig:mf}(b) are the same as those of the mean-field result shown in the main text.
In this case, we find two excitonic phases, EC1 and EC2, in addition to the uniform LS and HS phases.
The EC1 state is characterized by the uniform excitonic order parameters $\bar{\tau}^+$ and $\bar{\rho}^+$.
Since $\bar{n}_2$ is smaller than $\bar{n}_0$ and $\bar{n}_1$, this state is interpreted as condensation of single-excitons $\ket{1}$.
On the other hand, in the EC2 state, the staggered component of the order parameters continuously appears from the EC1 phase, indicating the second-order phase transition between EC1 and EC2.
Among them, $\bar{n}_1^-$ is much smaller than the others, which suggests that the solidification of bi-excitons emerges together with excitonic condensation.
Thus, the EC2 is regarded as an excitonic supersolid.
This phase is shrunk with increasing $V$ as shown in Supplementary Fig.~\ref{fig:mf}(c).
On the other hand, Supplementary Fig.~\ref{fig:mf}(a) indicates that the decrease of $V$ changes the EC2 to the classical ES state.
In this case, the phase transition between EC1 and ES is of the first order.
Therefore, we deduce that the fusion of the two single-excitons, caused by ${\cal H}_{\rm dual}$, plays a crucial role in inducing the EC2 state and the second-order transition to the EC1. 

\bl{
The phase diagrams on the plane of $V/E_{1}$ and $h/E_{1}$ are shown in Supplementary Figs.~\ref{fig:phase}(a)-\ref{fig:phase}(c) with the variations of $t/E_{1}=$ 0.03, 0.05, and 0.07, respectively.
The other parameters are chosen to be ($E_{1}$, $E_{2}$, $J_{11}$, $J_{02}$) =  (1.0, 0.7, 0.2, 0.05).
The phase diagrams indicate that the competition is in play between the classical interaction terms ${\cal H}_{\rm int}$ and the quantum terms ${\cal H}_{\rm trans}$, ${\cal H}_{\rm dual}$.
And also the competition is in play between the quantum terms ${\cal H}_{\rm trans}$ and ${\cal H}_{\rm dual}$.
They show that the EC1, EC2, and ES phases emerge between the LS and HS phases.
With increasing $V/E_{1}$, the quantum phases of EC1 and EC2 emerge with suppression of the classical solid ES.
With increasing $t/E_{1}$, EC1 becomes more stable as compared to EC2.
With $t/E_{1} = 0.07$ and $V/E_{1} = 0.17$, the transition from LS to EC1 and then to EC2 appears, which is the parameter set 1 presented in the main text and Supplementary Fig.~\ref{fig:phase}(c).
With $t/E_{1} = 0.07$ and $V/E_{1} = 0.13$, the transition from LS to ES and then to HS appears, which is the parameter set 2 presented in the main text and Supplementary Fig.~\ref{fig:phase}(c). 
}

\subsection{Relation to the experimental results}

We argue that the experimentally observed transitions from LS $\leftrightarrow$ $\beta$ $\leftrightarrow$ $\gamma$ possibly correspond to the successive phase transitions in the calculated results, $\rm{LS} \leftrightarrow$ EC1 $\leftrightarrow$ EC2, considering the experimentally observed features that two phases appear in high magnetic fields, which show continuous change of exciton density as a function of magnetic fields as shown in Supplementary Fig. 3g.
We also argue that the experimentally observed $\alpha$ phase originates in the ES phase in Supplementary Fig. \ref{fig:mf}a, considering the wide plateau of exciton density in ES and the $\alpha$ phase.

We tentatively compare the transition fields of the experiment and the calculation.
$E_1$, the energy of isolated IS state, is reported to be $\sim0.2$ eV with a CoO$_{6}$ cluster calculation in Ref. \cite{Haverkort}, which is assumed to be unity in the calculation.
Then, we obtain $h/E_1 = 0.1 \rightarrow 172.7$ T as a scale of the horizontal axis of Supplementary Figs. \ref{fig:mf}a-\ref{fig:mf}c, with the Zeeman energy of $E = g\mu_{\rm{B}}BS_z$.
We obtain the transition fields of (161 T, 193 T, 432 T
), (117 T, 263 T, 444 T), and (73 T, 307 T, 418 T, 485 T) for Supplementary Figs. \ref{fig:mf}a-\ref{fig:mf}c.
As an order estimation, the obtained transition fields well account for the experimental result.

The negative slope in the $\beta$ phase may result from the non-trivial spin-lattice coupling found in a high-resolution x-ray diffraction study, where the lattice volume of $\ket{1}$ is claimed to be smaller than a pair of neighboring site $\ket{0}$ and $\ket{2}$  \cite{Radaelli}.
Taking this into account, increasing the weight of $\ket{1, 1}$ will contract the lattice volume in EC1 with increasing magnetic fields.
On the other hand, increasing the weight of $\ket{0, 2}$ state will expand the lattice volume in EC2.

We note that the $\alpha$ phase may emerge with modified interaction parameters as compared to those for the $\beta$ and $\gamma$ phases, as ES appears with the variation of the HS-IS duality parameter $V$ suppressing the appearance of EC1 and EC2. 
The modification of the parameter should occur due to the strong lattice contraction, which enhances the localization nature of excitons, resulting in exciton superlattice formation in the $\alpha$ phase.
Such lattice contraction in the course of the $\beta$-$\alpha$ phase transition above 100 T is actually observed in Ref. \cite{IkedaPRL2020}.
Further microscopic understanding of the lattice change affecting the interaction parameters will be an interesting work in the future which will be accomplished using the state-of-the-art technique utilizing an x-ray free electron laser and a portable generator of ultrahigh magnetic fields \cite{IkedaAPL2022}.

\section*{Supplementary References}
\bibliography{lco_sup}